\documentclass[twocolumn,preprintnumbers,amsmath,amssymb]{revtex4}  

\usepackage{float}
\usepackage{graphicx}
\usepackage{dcolumn}
\usepackage{bm}

\begin{document}

\title{Searching for two-dimensional Weyl superconductors in heterostructures}

\author{Lei Hao$^{1,2}$ and C. S. Ting$^{1}$}
 \address{$^1$ Texas Center for Superconductivity and Department of Physics,
University of Houston, Houston, Texas 77204, USA  \\$^2$ Department of Physics, Southeast University, Nanjing 210096, China}

\date{\today}

\begin{abstract}
Two-dimensional Weyl superconductor is the most elusive member of a group of materials with Weyl fermions as low-energy excitations. Here, we propose to realize this state in a heterostructure consisting of thin films of half-metal and spin-singlet superconductor. In particular, for the $d$-wave case, a very robust two-dimensional Weyl superconductor (dWSC) is realized independent of the orientation of the spontaneous magnetization of the half-metal. The quasiparticle spectra of the dWSC show interesting evolution with the direction of the magnetization, featured by a series of Lifshitz transitions in the zero-energy contour of the quasiparticle spectrum. In addition, we find a transition between type-I and type-II Weyl nodes. This is also the first example of a two-dimensional type-II Weyl node in the presence of superconducting correlation. For a general magnetization orientation of the half-metal, the state is a combination of a superconducting component and a normal fluid component and is different from all known forms of pairings. The symmetries and topological properties of the system are analyzed. We also study the phases in the heterostructure with the half-metal replaced by a ferromagnetic metal with a partially spin-polarized Fermi surface.
\end{abstract}


\maketitle

\section{\label{sec:Introduction}introduction}

Searching for new materials with nontrivial topological properties is a forefront research field of condensed matter physics. Following the theoretical prediction and experimental discovery of two-dimensional and three-dimensional topological insulators \cite{kane05,bernevig06,fu07,moore07,roy09,zhang09,konig07,hsieh08,xia09}, the fully-gapped topological insulators and topological superconductors have been classified according to their symmetries \cite{hasan10,qi11,schnyder08,kitaev09,chenx13,mesaros13,lu16}. One focus of recent research is on a class of materials with discrete degeneracy points in the band structure \cite{wan11,burkov11,murakami07}. These Dirac and Weyl semimetals are analogies or generalizations of graphene, which is a prototypical two-dimensional (2D) Dirac semimetal \cite{wehling14}. The three-dimensional (3D) Weyl semimetals realize the chiral anomaly which results in several novel transport properties, including the anomalous Hall effect and the chiral magnetic effect \cite{yang11,hosur12,zyuzin12,liu13,hosur13}. The interest in 3D Weyl semimetal is soon extended to 2D Dirac and Weyl semimetals \cite{young12,young15,wang12,wang13,yang14nc,klinovaja16} and 3D Dirac and Weyl superconductors \cite{meng12,cho12,yang14,sato16}.

The 2D Weyl superconductor, which is also a member of this new family of materials characterized by the presence of Weyl fermions, has however till now received very little attention. This is partly because of the difficulty in realizing such a phase. On one hand, accidental band degeneracy is known to be very unlikely in 2D \cite{neumann29,blount85,balents11,hou13}. In a superconductor, at least one particle band and one hole band (which are related to each other by charge conjugation) cross each other and thus invert in a certain region of the wave vector space, which makes possible the occurrence of nodal points as the parts of the crossing curve between the particle and hole bands that are not gapped out. However, by opening a full gap the superconductor can usually gain more condensation energy and so naturally occurring nodal superconductors are rare. On the other hand, the realization of Weyl nodes requires non-degenerate Fermi surface. This requires the breaking of at least one of the time-reversal symmetry and the inversion symmetry \cite{hosur13}. While the former requires the persistence of magnetic order in the superconducting phase, the latter point to noncentrosymmetric materials. In bulk materials, the competition with other symmetry breaking tendencies makes the superconducting phase less favored as compared to the simple time-reversal invariant and inversion symmetric materials. Therefore, the 2D Weyl superconductor appears to be the most elusive member of the Weyl materials.

In this work, we propose that a 2D Weyl superconductor can be created in a very inclusive system, a heterostructure consisting of a half-metal (HM) with Rashba spin-orbital coupling (RSOC) and a spin-singlet superconductor. The combination of a single nondegenerate Fermi surface \cite{lee09,chung11}, the transmutation between spin-singlet and spin-triplet pairing by the RSOC \cite{gorkov01,zhang08,fu08,sato09,sau10,alicea10}, and the possible existence of nodes in the order parameter of the spin-singlet pairing, make the HM layer of the proposed heterostructure a unique platform to realize the 2D Weyl superconductor. An important tunability of the system, sketched in Fig.1, is the orientation of magnetization of the HM. Earlier, the heterostructure with an $s$-wave superconductor was found to give a 2D Weyl superconductivity phase (sWSC) when the spontaneous magnetization of the HM lies along a mirror symmetric direction within the film plane \cite{hao16}. In the heterostructure with a $d$-wave superconductor (such as the cuprates), on the other hand, a 2D Weyl superconductor (dWSC) is realized for an arbitrary magnetization orientation (the case with purely out-of-plane magnetization was also studied by Sato and Fujimoto in a slightly different setting \cite{sato10}, see also related works in Ref.\cite{wongetal}). The bulk and edge state spectra and their evolution with the magnetization direction show distinct differences from those for the sWSC. For example, besides the type-I Weyl nodes which are realized in sWSC, we also find type-II Weyl nodes in a dWSC. In Weyl semimetal, type-I and type-II Weyl nodes are defined separately as Weyl point with a point-like Fermi surface and Weyl point appearing at the contact of electron and hole pockets \cite{soluyanov15,wang16,chang16,xu15}. Here, we define type-I and type-II Weyl nodes in a superconductor analogously. Namely, for type-I Weyl node the constant energy surface (not necessarily the Fermi surface) crossing the Weyl point spans a single point, whereas for type-II Weyl node the constant energy surface crossing the Weyl point forms an electron-like and a hole-like quasiparticle `Fermi surface' and connect to each other at the Weyl point. The transition between type-I and type-II Weyl nodes is achieved in a dWSC by varying the direction of the spontaneous magnetization of the HM. In addition, for a general magnetization orientation, the state studied consists simultaneously of normal and superfluid components. In terms of the zero-energy quasiparticle spectrum, the varying magnetization orientation drives the system to go through a series of Lifshitz transitions, which is usually encountered in a metal in the normal phase \cite{lifshitz}. It is also shown to be qualitatively different from all known types of pairings, including the BCS pairing \cite{bcs}, the FFLO pairing \cite{ff,lo}, and the breached pairing state (Sarma phase) \cite{liu03,gubankova03,bedaque03,sarma63}.

The organization of this paper is as follows. In Sec. II, we define the model and parameters for the proposed heterostructure. In Sec. III, we report the energy spectra of the system. These include the energy spectra for bulk and a strip of the system, the spectral functions and density of states for the two edges of a strip, the clarification of the nature of proximity-induced superconducting state, and the Lifshitz transitions in the quasiparticle spectrum. The transition between type-I and type-II Weyl nodes is also studied in this section. (The analytical analyses are put to Appendix A.) In Sec. IV, we study the symmetry and topological properties of the system. In Sec. V, we show the energy spectra for the heterostructure with the HM replaced by an ordinary ferromagnetic material having a partially polarized Fermi surface. Finally, we summarize our results in Sec. VI.

\section{model}

\begin{figure}[!htb]\label{fig1} \centering
\includegraphics[width=7.5cm,height=3.0cm]{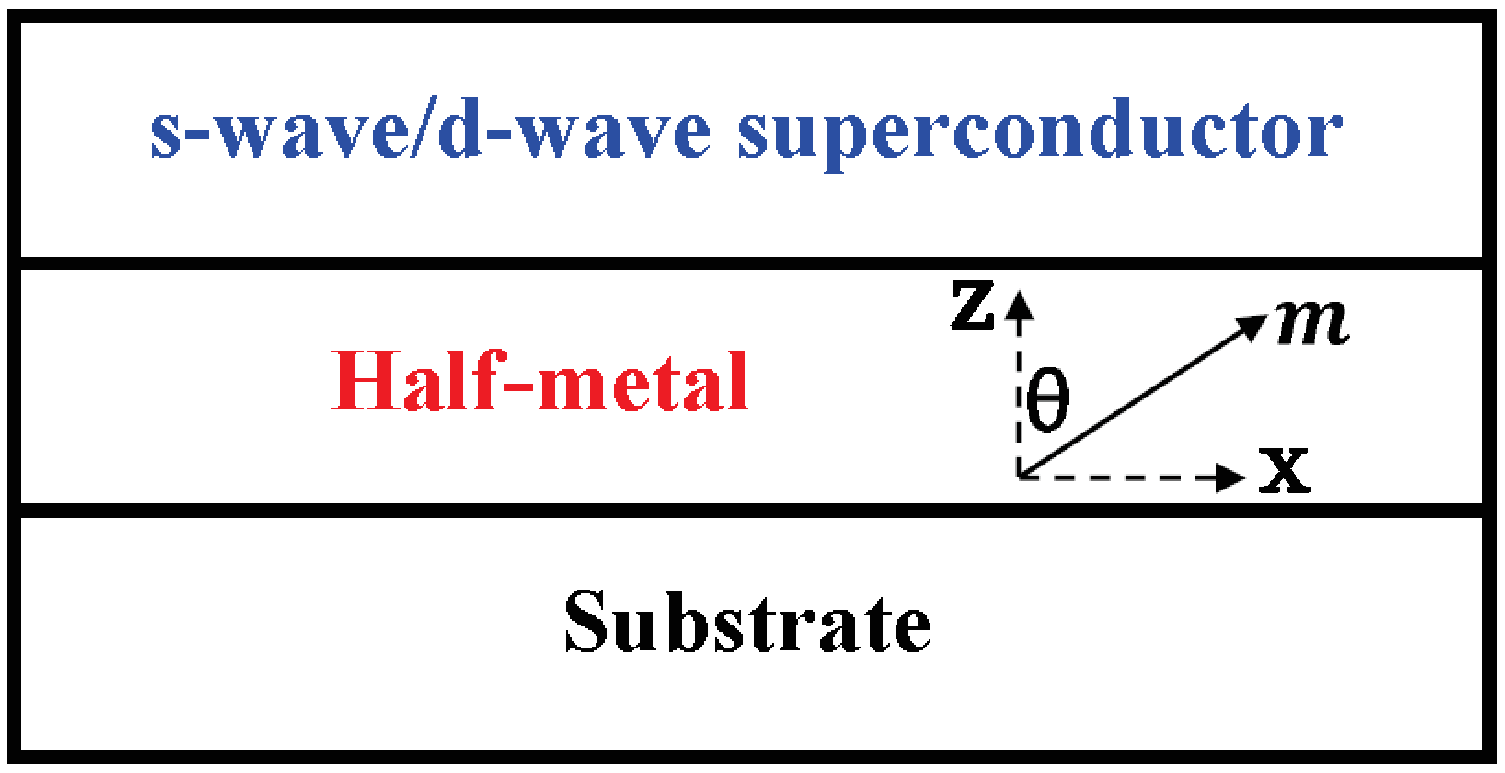}  \\
\caption{(Color online)Schematic drawing of a heterostructure consisting of a half-metal (HM) thin film sandwiched between an $s$-wave or $d$-wave spin-singlet superconductor and an insulating substrate, viewed laterally. $m$ and $\theta$ represent the magnitude and
direction of the exchange field (magnetization) in the HM, which is assumed to lie on the $xz$ plane.}
\end{figure}

Figure 1 is a schematic plot of the system that we will study. It consists of a HM (or, highly polarized ferromagnetic metal) thin film grown on an insulating substrate and capped with a spin-singlet superconductor with $s$-wave or $d$-wave pairing symmetry. To elucidate the principle more clearly, we consider the simplest model for the system. We describe the HM thin film by a one-orbital model defined on a square lattice. Perfect interfaces between the HM and the substrate and the spin-singlet superconductor are assumed, so that we will ignore the effect of impurity and disorder. Denoting the basis vector as
$\phi^{\dagger}_{\mathbf{k}}=[d^{\dagger}_{\mathbf{k}\uparrow},d^{\dagger}_{\mathbf{k}\downarrow}]$, the model Hamiltonian for the HM thin film, with a RSOC term induced by the formation of the
heterostructure, is $\hat{H}_{0}=\sum_{\mathbf{k}}\phi^{\dagger}_{\mathbf{k}}h_{0}(\mathbf{k})\phi_{\mathbf{k}}$, where \cite{sau10,chung11}
\begin{equation}
h_{0}(\mathbf{k})=\epsilon_{\mathbf{k}}\sigma_{0}+m_{x}\sigma_{1}+m_{z}\sigma_{3}+\lambda(\sin k_{x}\sigma_{2}-\sin k_{y}\sigma_{1}).
\end{equation}
$\sigma_{0}$ is the two-dimensional unit matrix, $\sigma_{i}$ ($i=1,2,3$) are Pauli matrices in the spin subspace. $\epsilon_{\mathbf{k}}=-2t(\cos k_{x}+\cos k_{y})-\mu$. $t$ is the hopping amplitude and we assume $t>0$, $\mu$ is the chemical potential, $\lambda$ is the strength of the RSOC. In writing down Eq.(1), we have assumed the magnetization to be fixed on the $xz$ plane. In terms of $m$ ($m>0$) and $\theta$ ($0\le\theta<2\pi$, with respect to the positive direction of the $z$ axis), which are separately the magnitude and polar angle of the magnetization (see Fig.(1)), we have $m_{x}=m\sin\theta$, $m_{z}=m\cos\theta$. To describe a HM with $\hat{H}_{0}$, we set $m$ to be of the same order of magnitude as $t$, and is much larger than the externally induced $\lambda$. The chemical potential is tuned to make sure that it crosses only the lower spin-split band of $h_{0}(\mathbf{k})$, which amounts to $\mu\simeq -4t$.

The proximity-induced superconductivity in the HM arising from coupling with an $s$-wave or $d$-wave spin-singlet superconductor is described by
\begin{equation}
\hat{H}_{p}=\frac{1}{2}\sum_{\mathbf{k}}\phi^{\dagger}_{\mathbf{k}}\underline{\Delta}(\mathbf{k})\phi^{\dagger}_{-\mathbf{k}}+\text{H.c.},
\end{equation}
where $\underline{\Delta}(\mathbf{k})=\Delta_{0}(\mathbf{k})i\sigma_{2}$ \cite{fu08,sau10,hao16}. $\Delta_{0}(\mathbf{k})=\Delta_{0}\eta(\mathbf{k})$ is a multiplication of the pairing amplitude $\Delta_{0}$ and a symmetry factor $\eta(\mathbf{k})$. For the sake of simplicity, we take $\eta(\mathbf{k})=1$ for $s$-wave superconductor and $\eta(\mathbf{k})=\cos k_{x}-\cos k_{y}$ or $\eta(\mathbf{k})=\sin k_{x}\sin k_{y}$ for $d$-wave superconductor \cite{fu08,sau10,sato10}. The two latter cases can be realized by growing the same $d$-wave superconductor (e.g., cuprates) on the top of the HM film along two special directions, respectively. The $x$-axis, which is the projection of the magnetization plane ($xz$) onto the plane of the films ($xy$) in the heterostructure, is along the antinodal direction of the $d$-wave pairing gap for $\eta(\mathbf{k})=\cos k_{x}-\cos k_{y}$ and along the nodal direction of the $d$-wave pairing gap for $\eta(\mathbf{k})=\sin k_{x}\sin k_{y}$. The heterostructure formed by growing the $d$-wave superconductor along more general (less symmetric) directions, while possible in principle, can be understood in terms of a combination of the results for these two special cases. In the Nambu basis, $\varphi^{\dagger}_{\mathbf{k}}=[\phi^{\dagger}_{\mathbf{k}},\phi^{\text{T}}_{-\mathbf{k}}]$, the full model is written as
$\hat{H}=\frac{1}{2}\sum_{\mathbf{k}}\varphi^{\dagger}_{\mathbf{k}}h(\mathbf{k})\varphi_{\mathbf{k}}$, where the Bogoliubov-de Gennes (BdG) Hamiltonian with proximity-induced pairing term and RSOC is
\begin{eqnarray}
h(\mathbf{k})&=&\epsilon_{\mathbf{k}}\tau_{3}\sigma_{0}+m_{x}\tau_{3}\sigma_{1}+m_{z}\tau_{3}\sigma_{3}    \notag \\
&&+\lambda(\sin k_{x}\tau_{3}\sigma_{2}-\sin k_{y}\tau_{0}\sigma_{1})-\Delta_{0}\eta(\mathbf{k})\tau_{2}\sigma_{2}.
\end{eqnarray}
$\tau_{i}$ ($i=1,2,3$) are Pauli matrices in the Nambu space. Diagonalizing Eq.(3) gives the four quasiparticle bands $E_{n}(\mathbf{k})$, with $n$=$-2$, $-1$, $1$, $2$ and larger $n$ means higher energy.

In the HM thin film, only the spin-polarized band crossing the chemical potential is important to the low-energy properties of the system. To facilitate the analysis, we construct the low-energy effective model by projecting to the subspace containing only information of the band contributing to the Fermi surface. First, we diagonalize the model defined in Eq.(1) by a unitary transformation $U^{\dagger}(\mathbf{k})h_{0}(\mathbf{k})U(\mathbf{k})=h_{d}(\mathbf{k})$. $h_{d}(\mathbf{k})$ is a diagonal matrix with the two diagonal elements storing the eigenvalues of $h_{0}(\mathbf{k})$,
\begin{eqnarray}
E_{\alpha}(\mathbf{k})&=&\epsilon_{\mathbf{k}}+\alpha\sqrt{m_{z}^{2}+(m_x-\lambda\sin k_{y})^{2}+\lambda^{2}\sin^{2} k_{x}} \notag \\
&=&\epsilon_{\mathbf{k}}+\alpha\tilde{E}(\mathbf{k}),
\end{eqnarray}
where $\alpha=\pm$. Notice that, $E_{\alpha}(\mathbf{k})$ ($\alpha=\pm$) are the electronic energy bands in the normal state (i.e., without the proximity-induced pairing term) and should be distinguished from the quasiparticle bands $E_{n}(\mathbf{k})$ ($n=$-2, -1, 1, 2) defined as eigenvalues of Eq.(3). The unitary matrix $U(\mathbf{k})$ containing the eigenvectors of $h_{0}(\mathbf{k})$ is written as
\begin{equation}
U(\mathbf{k})=\begin{pmatrix} u_{+}(\mathbf{k}) & u_{-}(\mathbf{k}) \\
v_{+}(\mathbf{k}) & v_{-}(\mathbf{k})
\end{pmatrix}.
\end{equation}
The normalized eigenvectors are chosen as
\begin{equation}
\begin{pmatrix} u_{\alpha}(\mathbf{k}) \\
v_{\alpha}(\mathbf{k})\end{pmatrix}=\frac{1}{D_{\alpha}(\mathbf{k})}\begin{pmatrix}
m_{x}-\lambda\sin k_{y}-i\lambda\sin k_{x} \\
\alpha\tilde{E}(\mathbf{k})-m_{z} \end{pmatrix},
\end{equation}
where $\alpha=\pm$ and $D_{\alpha}(\mathbf{k})=\sqrt{2\tilde{E}(\mathbf{k})(\tilde{E}(\mathbf{k})-\alpha m_{z})}$. $\tilde{E}(\mathbf{k})$ is defined in Eq.(4). In the eigenbasis, the pairing term is rewritten as
\begin{equation}
\tilde{\underline{\Delta}}(\mathbf{k})=U^{\dagger}(\mathbf{k})\underline{\Delta}(\mathbf{k})U^{\ast}(-\mathbf{k}).
\end{equation}
For $\mu\sim-4t$ that we will focus on, only the $E_{-}(\mathbf{k})$ band contributes to the low-energy properties. We thus retain only this band and the pairing term within this band to construct the low-energy effective model. In the reduced Nambu space defined by the basis $\tilde{\varphi}^{\dagger}_{\mathbf{k}}=[d^{\dagger}_{\mathbf{k},-},d_{-\mathbf{k},-}]$, the low-energy effective model is written as
\begin{eqnarray}
\tilde{H}&=&\frac{1}{2}\sum\limits_{\mathbf{k}}\tilde{\varphi}^{\dagger}_{\mathbf{k}}\begin{pmatrix} E_{-}(\mathbf{k}) &  \tilde{\Delta}_{--}(\mathbf{k}) \\
\tilde{\Delta}^{\ast}_{--}(\mathbf{k}) & -E_{-}(-\mathbf{k})
\end{pmatrix}\tilde{\varphi}_{\mathbf{k}}   \notag \\
&=&\frac{1}{2}\sum\limits_{\mathbf{k}}\tilde{\varphi}^{\dagger}_{\mathbf{k}} \tilde{h}(\mathbf{k}) \tilde{\varphi}_{\mathbf{k}}.
\end{eqnarray}
The intra-band pairing amplitude is
\begin{eqnarray}
\tilde{\Delta}_{--}(\mathbf{k})&=&\Delta_{0}\eta(\mathbf{k})[u^{\ast}_{-}(\mathbf{k})v^{\ast}_{-}(-\mathbf{k})-u^{\ast}_{-}(-\mathbf{k})v^{\ast}_{-}(\mathbf{k})]  \notag \\
&=&\frac{2\Delta_{0}\eta(\mathbf{k})}{D_{-}(\mathbf{k})D_{-}(-\mathbf{k})}[\lambda^{\ast}_{\mathbf{k}}\tilde{m}_{z}(\mathbf{k})+\tilde{m}_{x}(\mathbf{k})]  \notag \\ &\equiv&\Delta_{0}\eta(\mathbf{k})f(\mathbf{k}),
\end{eqnarray}
where we have introduced the abbreviations $\lambda_{\mathbf{k}}=\lambda(\sin k_{y}+i\sin k_{x})$, $\tilde{m}_{z}(\mathbf{k})=m_{z}-\frac{1}{2}(\tilde{E}(\mathbf{k})+\tilde{E}(-\mathbf{k}))$, and $\tilde{m}_{x}(\mathbf{k})=m_{x}(\tilde{E}(-\mathbf{k})-\tilde{E}(\mathbf{k}))$. The effective paring is therefore determined by two symmetry factors, $\eta(\mathbf{k})$ and $f(\mathbf{k})$. The eigenstates of $\tilde{h}(\mathbf{k})$ will be used as approximations for the two low-energy quasiparticle bands, $E_{-1}(\mathbf{k})$ and $E_{1}(\mathbf{k})$, in the following analytical analysis.

\section{energy spectrum}

For the heterostructure with $s$-wave superconductor and in-plane exchange field ($\theta=\pi/2$ or $\theta=3\pi/2$), the proximity-induced superconducting phase in the HM thin film is an sWSC with a single pair of Weyl nodes located on the mirror invariant line of the 2D Brillouin zone (BZ) \cite{hao16}. It has also been verified that, the sWSC is a direct result of an emergent mirror reflection symmetry that exists only for an in-plane exchange field \cite{hao16}.

The heterostructure with a $d$-wave superconductor, however, supports a dWSC phase independent of the $\theta$ angle of the exchange field. This is understandable from the intraband pairing amplitude in Eq.(9). Namely, because of the presence of the $\eta({\mathbf{k}})$ factor, there are always point nodes inherited from those of the $d$-wave superconductor. The remaining wave-vector dependence of the effective pairing amplitude, contained in $f(\mathbf{k})$, is the same as that in the heterostructure with $s$-wave superconductor. In particular, $f(k_{x}=0,\theta=\frac{\pi}{2})=0$ \cite{hao16}. From the positions of the nodes resulting from the $\eta(\mathbf{k})$ factor and from the $f(\mathbf{k})$ factor, we know that as $\theta$ varies the number of pairs of Weyl nodes vary between two and three for $\eta(\mathbf{k})=\cos k_{x}-\cos k_{y}$ and remains to be two in the whole range of $\theta$ for $\eta(\mathbf{k})=\sin k_{x}\sin k_{y}$. We study the energy spectra of the two cases separately in what follows.

\subsection{Quasipaticle spectrum for $\eta(\mathbf{k})=\cos k_{x}-\cos k_{y}$}

Two kinds of topological phase transitions were identified in the heterostructure with $s$-wave superconductor, as we vary $\theta$ \cite{hao16}. One is between a fully gapped phase and a bulk gapless phase with ubiquitous local gap. Another is between the bulk gapless phase with ubiquitous local gap and the sWSC phase with a pair of Weyl nodes connecting the two low-energy quasiparticle bands $E_{-1}(\mathbf{k})$ and $E_{1}(\mathbf{k})$. For the heterostructure with $d$-wave superconductor, the persistence of the Weyl nodes for all $\theta$ angles makes exact analogies of the corresponding phase transitions nonexistent. However, we can still define several critical angles around which to expect some qualitative changes in the quasiparticle energy spectrum for the system in bulk and strip configurations. In this respect, for $\eta(\mathbf{k})=\cos k_{x}-\cos k_{y}$, we can define three critical angles in the range $\theta\in[0,\pi/2]$ as follows.

The critical angles are defined in terms of the quasiparticle energies of the four Weyl nodes and those for two special Fermi points. For the model parameters of interest to us ($|\lambda/m|\ll1$), the positions of the reference wave vectors keep almost unchanged as $\theta$ varies. The four Weyl nodes are at $(\pm1,\pm1)k_{x0}$ with $k_{x0}>0$ determined by
\begin{equation}
\cos k_{x0}=\frac{-2\mu t}{8t^{2}+\lambda^{2}}[1-\sqrt{1-\frac{(8t^{2}+\lambda^{2})(\mu^{2}-m^{2}-2\lambda^{2})}{8\mu^{2}t^{2}}}].
\end{equation}
The two special Fermi points are chosen as $(0,\pm1)k_{y0}$, with $k_{y0}>0$ determined by
\begin{equation}
\cos k_{y0}=\frac{2mt}{\lambda^{2}}[1-\sqrt{1+\frac{\lambda^{2}(2m^{2}+\lambda^{2}+2m\mu+4mt)}{4m^{2}t^{2}}}].
\end{equation}
The critical angle $\theta_{c1}$ is determined by the condition $E_{1}(k_{x0},k_{x0})=E_{-1}(k_{x0},k_{x0})=E_{\text{sgn}(\lambda)1}(0,-k_{y0})$, where $\text{sgn}(x)$ is the sign function. The critical angle $\theta_{c2}$ is determined by the condition $E_{-\text{sgn}(\lambda)1}(0,k_{y0})=E_{\text{sgn}(\lambda)1}(0,-k_{y0})$. Making use of the low-energy effective model defined by Eqs.(8) and (9), $\theta_{c1}$ and $\theta_{c2}$ are determined approximately by
\begin{equation}
|\tan\theta_{c1}|=\frac{|\Delta_{0}|}{m}(1-\cos k_{y0})\frac{|\sin k_{y0}|}{2|\sin k_{x0}|},
\end{equation}
and
\begin{equation}
|\tan\theta_{c2}|=\frac{|\Delta_{0}|}{m}(1-\cos k_{y0}).
\end{equation}
The third critical angle is $\theta_{c3}=\pi/2$, at which the number of Weyl nodes changes from two pairs to three pairs. For small $|\lambda|/m$, $(0,\pm1)k_{y0}$ gives the approximate positions of the two additional Weyl nodes.

We proceed to show the numerical results for the energy spectra of the dWSC in both bulk and strip configurations, for several typical values of $\theta\in[0,\pi/2]$. The bulk energy spectra are shown along several lines with fixed values of $k_{x}\ge0$, since the bulk quasiparticle spectrum is symmetric with respect to $k_{x}$. We focus on a typical set of parameters as $t=1$, $m=t$, $\mu=-4.6t$, $\lambda=0.2t$, and $\Delta_{0}=0.5t$. A relatively large value of $\Delta_{0}$ is adopted to see the qualitative features of the spectra more clearly.

\begin{figure}[!htb]\label{fig2} \centering
\includegraphics[width=8.5cm,height=15.0cm]{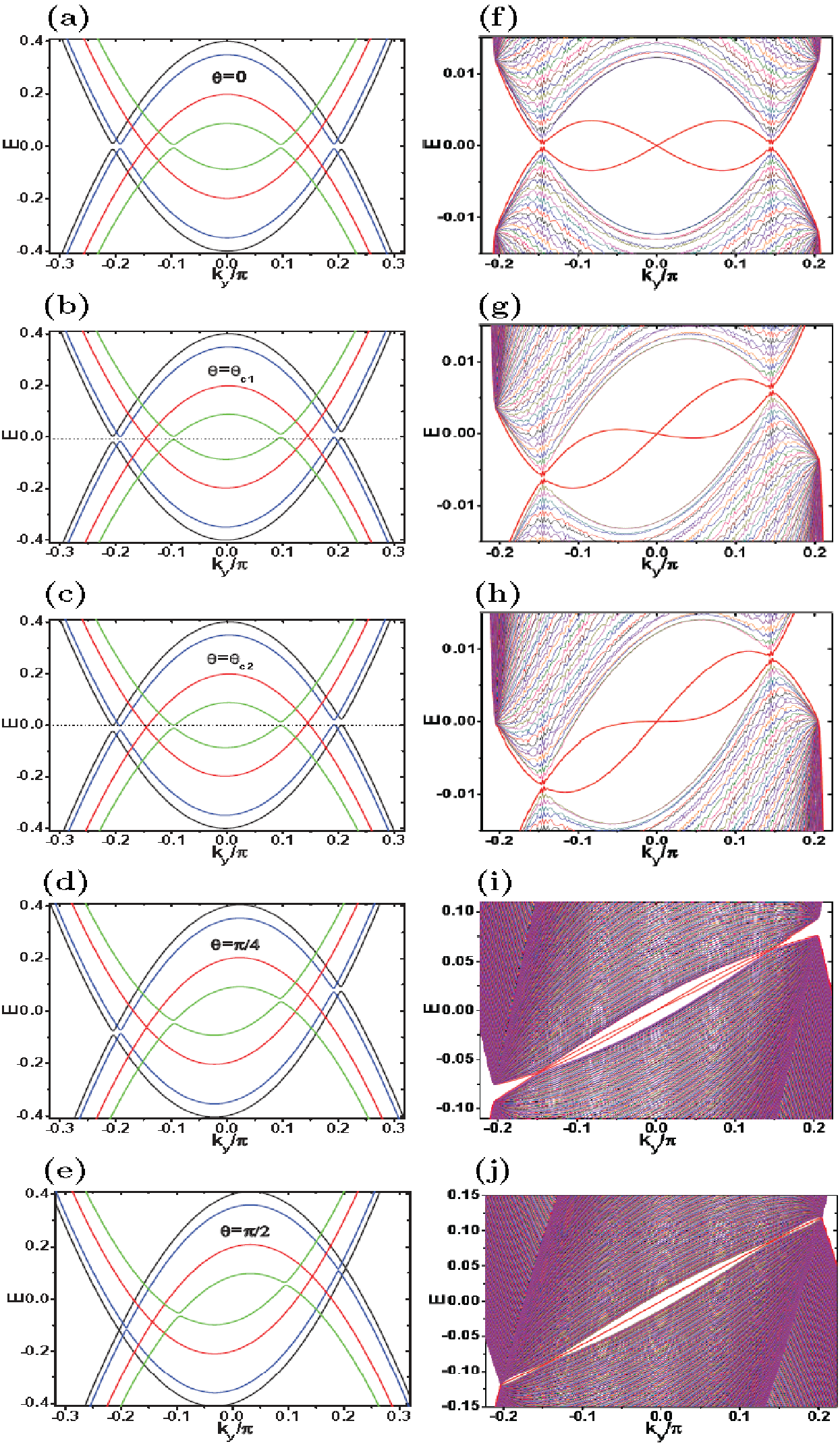}
\caption{(Color online)Bulk (a, b, c, d, e) and edge (f, g, h, i, j) state spectra of the dWSC with $\eta(\mathbf{k})=\cos k_{x}-\cos k_{y}$, for a typical set of parameters, $m=t=1$, $\lambda=0.2t$, $\Delta_{0}=0.5t$, $\mu=-4.6t$. For the bulk system, the dispersions of the $E_{-1}(\mathbf{k})$ and $E_{1}(\mathbf{k})$ bands are shown along four lines in the 2D BZ with fixed $k_{x}$ values at 0 (black lines), $\frac{1}{2}k_{x0}$ (blue lines), $k_{x0}$ (red lines), and $\frac{5}{4}k_{x0}$ (green lines). $k_{x0}>0$ is defined by Eq.(10). The edge state spectra correspond to a strip with 1500 unit cells along the $x$ direction. $\theta=0$ for (a) and (f), $\theta=\theta_{c1}$ for (b) and (g), $\theta=\theta_{c2}$ for (c) and (h), $\theta=0.25\pi$ for (d) and (i), $\theta=0.5\pi$ for (e) and (j).}
\end{figure}

Figure 2 shows the quasiparticle energy spectra for $\eta(\mathbf{k})=\cos k_{x}-\cos k_{y}$ in both the bulk [2(a) to 2(e)] and strip [2(f) to 2(j)] configurations. Five values of $\theta$, including $0$ [2(a) and 2(f)], $\theta_{c1}$ [2(b) and 2(g)], $\theta_{c2}$ [2(c) and 2(h)], $\frac{\pi}{4}$ [2(d) and 2(i)], and $\frac{\pi}{2}$ [2(e) and 2(j)] are considered. For the bulk spectra, we plot for each $\theta$ the quasiparticle bands $E_{-1}(\mathbf{k})$ and $E_{1}(\mathbf{k})$ along four lines in the 2D BZ with correspondingly fixed values of $k_{x}$, which are $0$, $\frac{1}{2}k_{x0}$, $k_{x0}$, and $\frac{3}{2}k_{x0}$. From Fig.2(b) and Fig.2(c), the critical angles $\theta_{c1}$ and $\theta_{c2}$ determined by Eq.(12) and Eq.(13) are very close to the exact values according to their definitions. As $\theta$ varies from $0$ to $\frac{\pi}{2}$, the positions of the Weyl nodes determined by the symmetry factor $\eta(\mathbf{k})=\cos k_{x}-\cos k_{y}$ keep almost unchanged. At $\theta=\pi/2$, it is clear from Fig.2(e) that, besides the four Weyl nodes determined by $\eta(\mathbf{k})=0$ that are always present, two additional nodes emerges along $k_{x}=0$ as a result of $f(k_{x}=0,\theta=\frac{\pi}{2})=0$ \cite{hao16}.

In the strip configuration (with two edges parallel to the $y$-axis), a pair of Weyl nodes with the same $k_{y}$ (equal to $k_{x0}$ or $-k_{x0}$) and energy project to the same point of the edge BZ. The breaking of translational symmetry along the $x$ direction gives a finite coupling between the two Weyl nodes. It is thus natural to expect that the Weyl nodes will be gapped out for a strip of finite width. Besides, it is known that there are edge modes for $\eta(\mathbf{k})=1$. It is interesting to see whether the edge states persists and how they are modified by the ubiquitously existing Weyl nodes. These questions are answered partly by the results in Figs. 2(f) to 2(j) for strips with 1500 unit cells along the $x$ direction. Clearly, a gap opens at the Weyl nodes. The edge modes persist but the dispersion is changed from the crossing-like structure to a shape similar to an $\infty$ symbol but then connected again to the edge modes beyond the projection of the Weyl nodes \cite{hao16}. Interestingly, the two edge modes are nondegenerate. This means that there are no additional edge modes associated with the Weyl nodes arising from the symmetry factor $\eta(\mathbf{k})$. For $\theta=\frac{\pi}{2}$, a third pair of Weyl nodes along $k_{x}=0$ emerges. Same as the case for the sWSC \cite{hao16}, the edge modes become Majorana Fermi lines connecting the two new Weyl nodes. In this case, since the edge modes pass through the projection of the two pairs of bulk Weyl nodes resulting from $\eta(\mathbf{k})=0$, the bulk nodes extending throughout the bulk of the system act as channels coupling the edge modes. Therefore, close to the projection of the bulk Weyl nodes at the edge BZ, the edge modes are re-hybridized.

\subsection{Quasipaticle spectrum for $\eta(\mathbf{k})=\sin k_{x}\sin k_{y}$}

For $\eta(\mathbf{k})=\sin k_{x}\sin k_{y}$, there are always two pairs of Weyl nodes for an arbitrary $\theta\in[0,2\pi)$. One pair is along $k_{x}=0$ and the other pair is along $k_{y}=0$. Since Weyl nodes exist along both of the two special high symmetry lines, there are no simple analogies to $\theta_{c1}$ and $\theta_{c2}$ defined for $\eta(\mathbf{k})=\cos k_{x}-\cos k_{y}$. Nevertheless, we can still define two critical angles in a manner similar to the above discussions. For $\lambda>0$ that we focus on in this work, the three special reference wave vectors for defining the two critical angles are chosen as $(0,k_{y0})$, $(k_{x0},k_{x0})$, and $(k_{x0},-k_{x0})$. $k_{x0}>0$ and $k_{y0}>0$ are determined by Eq.(10) and Eq.(11), respectively. At the first critical angle $\theta=\theta'_{c1}$, we have $E_{-1}(0,k_{y0})=E_{1}(0,k_{y0})=E_{1}(k_{x0},-k_{x0})$. At the second critical angle $\theta=\theta'_{c2}$, we have $E_{-1}(k_{x0},k_{x0})=E_{1}(k_{x0},-k_{x0})$. There are no simple analytical expressions for $\theta'_{c1}$ and $\theta'_{c2}$. We obtain them numerically by the definition.

The energy spectra of the system in the configurations of a bulk material and a strip are shown in Figure 3. The parameters as chosen as $t=1$, $m=t$, $\mu=-4.6t$, $\lambda=0.2t$, and $\Delta_{0}=0.5t$. We show the results for five values of $\theta$, including $0$ [3(a) and 3(f)], $\theta'_{c1}$ [3(b) and 3(g)], $\theta'_{c2}$ [3(c) and 3(h)], $\frac{\pi}{4}$ [3(d) and 3(i)], and $\frac{\pi}{2}$ [3(e) and 3(j)]. For the bulk spectra, we plot for each $\theta$ the quasiparticle bands $E_{-1}(\mathbf{k})$ and $E_{1}(\mathbf{k})$ along three lines in the 2D BZ with correspondingly fixed values of $k_{x}$, which are $0$, $k_{x0}$, and $k_{xc}$. $k_{x0}>0$ is still determined by Eq.(10). $k_{xc}>0$ is determined by
\begin{equation}
\cos k_{xc}=\frac{-2\tilde{\mu}t}{4t^{2}+\lambda^{2}}[1-\sqrt{1-\frac{(4t^{2}+\lambda^{2})(\tilde{\mu}^{2}-m^{2}-\lambda^{2})}{4\tilde{\mu}^{2}t^{2}}}],
\end{equation}
where $\tilde{\mu}=\mu+2t$. The persistence of two pairs of bulk Weyl nodes and their locations are confirmed by the results presented in Figs.3(a) to 3(e). A qualitative difference between the present case and those for the sWSC and the dWSC with $\eta(\mathbf{k})=\cos k_{x}-\cos k_{y}$ is, however, clear from the dispersions along lines with $k_{x}=k_{xc}$ on Figs.3(d) and 3(e). Namely, the Weyl nodes located at $(\pm k_{xc},0)$ change from the first type (type-I) to the second type (type-II) as $\theta$ increases and becomes larger than a critical value. This change in the character of the Weyl nodes has not been noticed in either the sWSC or the dWSC with $\eta(\mathbf{k})=\cos k_{x}-\cos k_{y}$. The pair of nodes located at $(0,\pm k_{y0})$, on the other hand, keeps to be type-I as $\theta$ changes. From Appendix A with more detailed analyses, the Weyl nodes at $(0,\pm k_{y0})$ are the only Weyl nodes in the present heterostructure (Fig.1) that can become type-II. This is the first case where type-II Weyl nodes are predicted in a 2D system and in the presence of superconducting correlation. It is also the first example of type-II Weyl nodes in all 2D systems, as far as we know.

\begin{figure}[!htb]\label{fig3} \centering
\includegraphics[width=8.5cm,height=15.0cm]{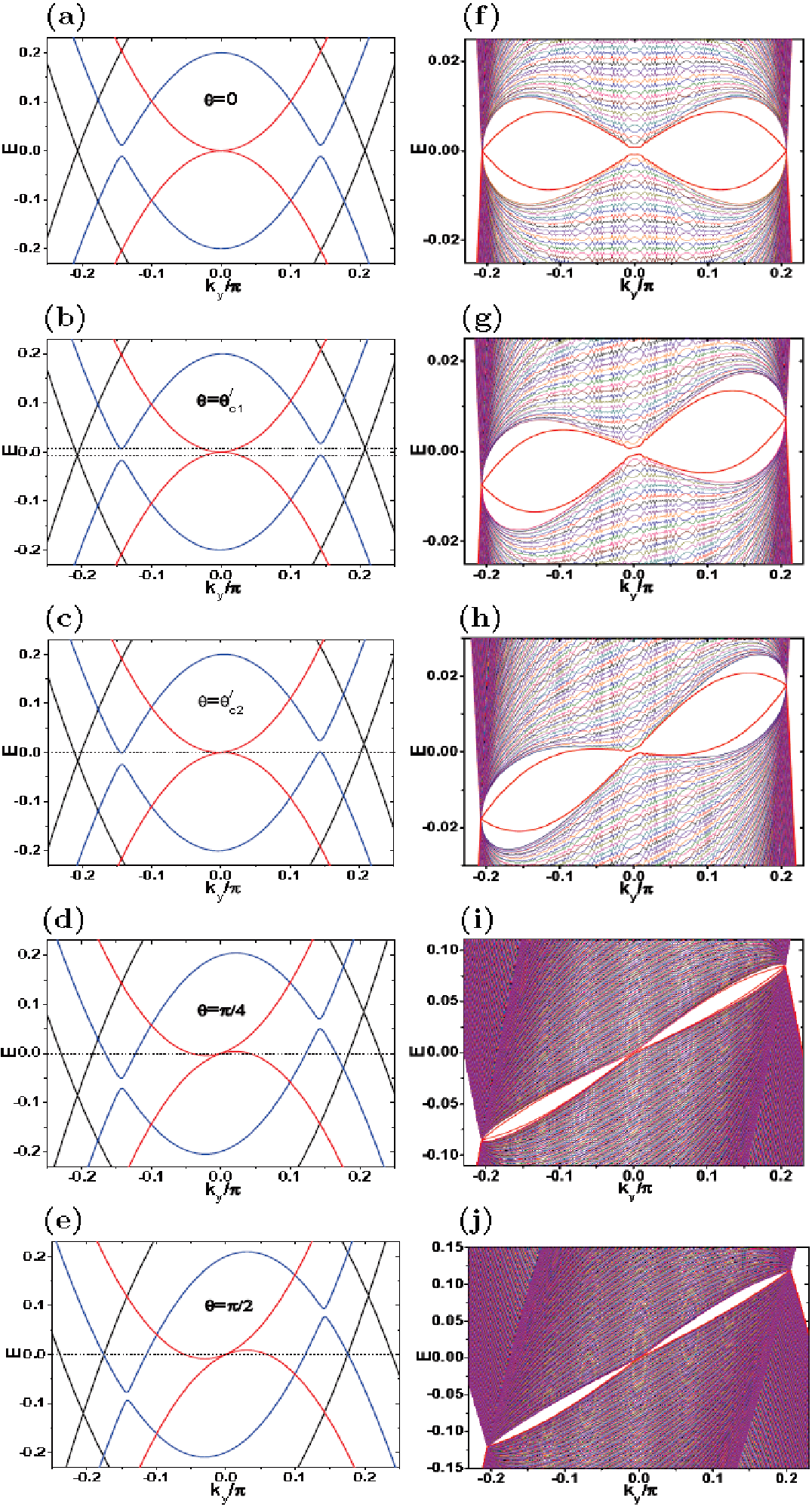}
\caption{(Color online)Bulk (a, b, c, d, e) and edge (f, g, h, i, j) state spectra of the dWSC with $\eta(\mathbf{k})=\sin k_{x}\sin k_{y}$, for a typical set of parameters, $m=t=1$, $\lambda=0.2t$, $\Delta_{0}=0.5t$, $\mu=-4.6t$. For the bulk system, the dispersions of the $E_{-1}(\mathbf{k})$ and $E_{1}(\mathbf{k})$ bands are shown along three lines in the 2D BZ with fixed $k_{x}$ values at 0 (black lines), $k_{x0}$ (blue lines), and $k_{xc}$ (red lines). The edge state spectra correspond to a strip with 1500 unit cells along the $x$ direction. $\theta=0$ for (a) and (f), $\theta=\theta'_{c1}$ for (b) and (g), $\theta=\theta'_{c2}$ for (c) and (h), $\theta=0.25\pi$ for (d) and (i), $\theta=0.5\pi$ for (e) and (j).}
\end{figure}

The energy spectra for a strip are shown in Figs.3(f) to 3(j), for five typical angles defined above. While the two Weyl nodes at $(\pm k_{xc},0)$ still project to the same point (i.e., $k_{y}=0$) of the edge BZ, the two Weyl nodes at $(0,\pm k_{y0})$ remain well separated from each other. The broken translational invariance along the $x$ direction couples the two Weyl nodes at $(\pm k_{xc},0)$ and opens a gap in the nodes, similar to the situation for $\eta(\mathbf{k})=\cos k_{x}-\cos k_{y}$. The edge modes also hybridize and open a gap at $k_{y}=0$ which is where the two Weyl nodes at $(\pm k_{xc},0)$ projects.

\subsection{Experimental features of the edge state spectra}

\begin{figure}[!htb]\label{fig4} \centering
\includegraphics[width=8.5cm,height=15.0cm]{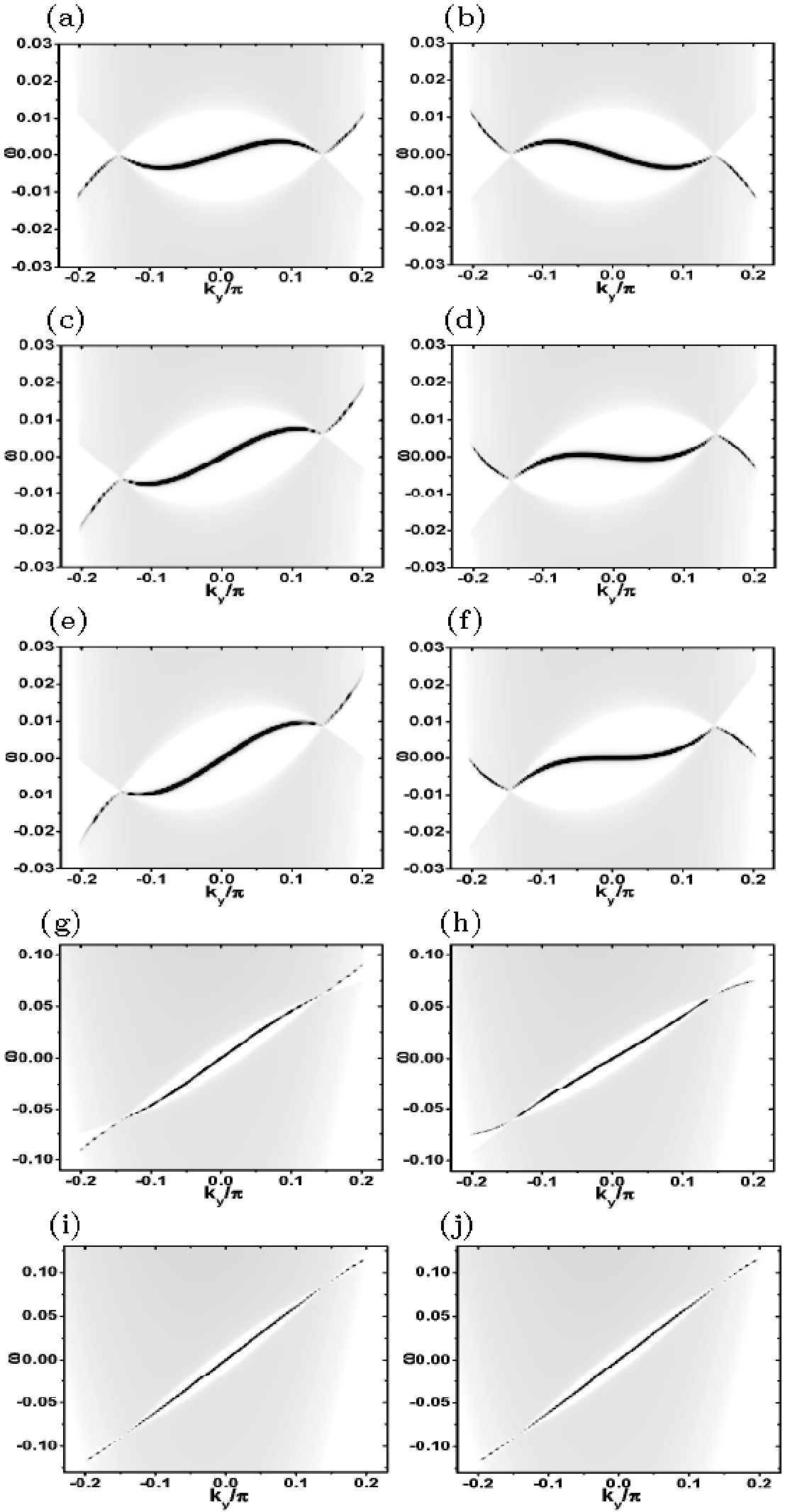}
\caption{The spectral functions for the left (a, c, e, g, i) and right (b, d, f, h, j) edges of a strip of the heterostructure with $\eta(\mathbf{k})=\cos k_{x}-\cos k_{y}$. Five values of $\theta$ are considered, including $\theta=0$ for (a) and (b), $\theta=\theta_{c1}$ for (c) and (d), $\theta=\theta_{c2}$ for (e) and (f), $\theta=0.25\pi$ for (g) and (h), $\theta=0.5\pi$ for (i) and (j). The parameters are taken as $m=t=1$, $\lambda=0.2t$, $\Delta_{0}=0.5t$, $\mu=-4.6t$. The darker the color, the larger the spectral weight.}
\end{figure}

\begin{figure}[!htb]\label{fig5} \centering
\includegraphics[width=8.5cm,height=15.0cm]{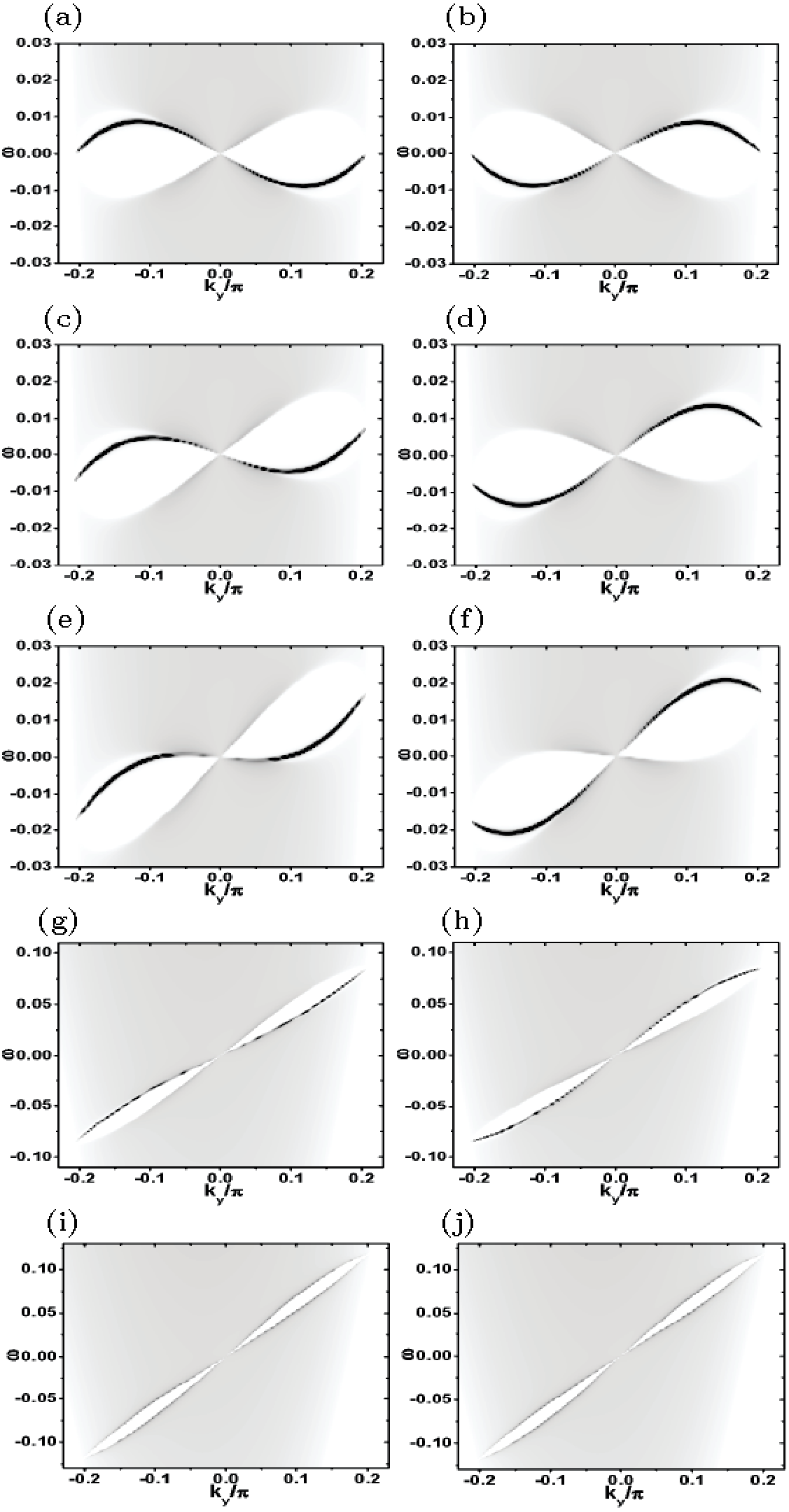}
\caption{The spectral functions for the left (a, c, e, g, i) and right (b, d, f, h, j) edges of a strip of the heterostructure with $\eta(\mathbf{k})=\sin k_{x}\sin k_{y}$. Five values of $\theta$ are considered, including $\theta=0$ for (a) and (b), $\theta=\theta'_{c1}$ for (c) and (d), $\theta=\theta'_{c2}$ for (e) and (f), $\theta=0.25\pi$ for (g) and (h), $\theta=0.5\pi$ for (i) and (j). The parameters are taken as $m=t=1$, $\lambda=0.2t$, $\Delta_{0}=0.5t$, $\mu=-4.6t$. The darker the color, the larger the spectral weight.}
\end{figure}

The energy spectra of the edge states can be probed by angle-resolved photoemission spectra (ARPES) \cite{arpes} and scanning tunneling spectroscopy (STS) \cite{sts}. Instead of the quasiparticle energy spectrum, the experimentally relevant quantity is the spectral function along the edge of a strip, which is defined as the imaginary part of the retarded Green's function for the edge layer of the strip \cite{hao16}. We have calculated the retarded Green's functions $G(n_{x},k_{y})$ for the left ($n_{x}=1$) and right ($n_{x}=N_{x}$) edges of a strip with $N_{x}$ layers, in terms of standard iterative Green's function method.\cite{hao11,hao16} As shown in Figs.4 and 5 are results separately for $\eta(\mathbf{k})=\cos k_{x}-\cos k_{y}$ and $\eta(\mathbf{k})=\sin k_{x}\sin k_{y}$, corresponding to five $\theta$ values taken separately in Figs.2 and 3. While the correspondences between the spectral functions in Figs.4 and 5 and the quasiparticle energy spectra in Figs.2 and 3 are very clear, the spectral functions contain important new information on how the edge states are distributed on the two edges. In addition, the Green's functions obtained in terms of the iterative Green's function method can be regarded as corresponding to a very wide strip ($N_{x}\rightarrow\infty$), there is thus no finite-size effect in the results. As a result, there is no gap opening in the spectral function at the Weyl nodes at $k_{y}=\pm k_{x0}$ for $\eta(\mathbf{k})=\cos k_{x}-\cos k_{y}$ and at $k_{y}=0$ for $\eta(\mathbf{k})=\sin k_{x}\sin k_{y}$.

\begin{figure}[!htb]\label{fig6}
\centering
\includegraphics[width=8.5cm,height=15.0cm]{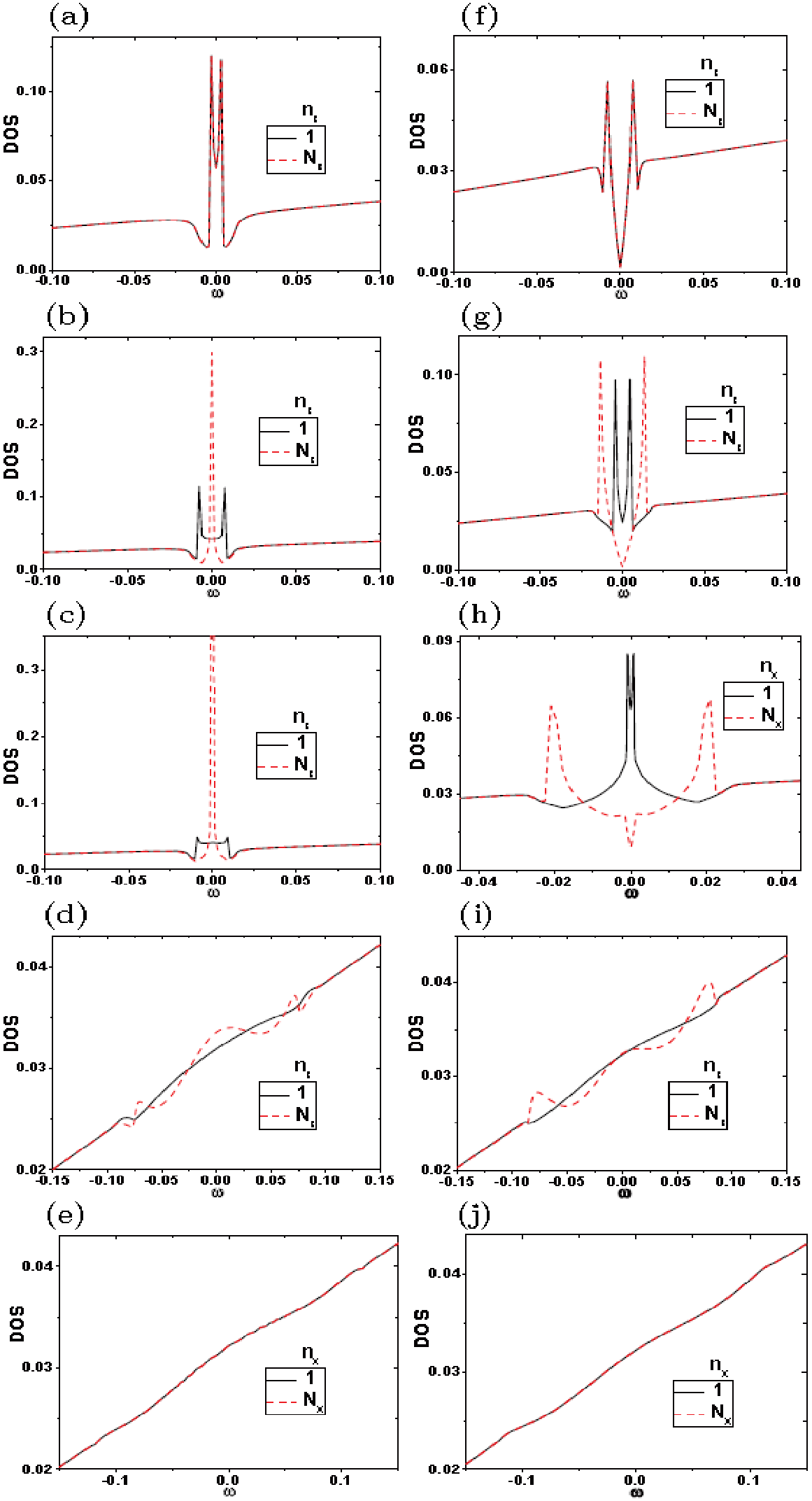}
\caption{(Color online) Density of states (DOS) on the two edges of a strip for $\eta(\mathbf{k})=\cos k_{x}-\cos k_{y}$ (a, b, c, d, e) and for $\eta(\mathbf{k})=\sin k_{x}\sin k_{y}$ (f, g, h, i, j). Results for the left and right edges are assigned respectively with a label of $n_{x}=1$ and $n_{x}=N_{x}$. In the calculation, the width of the strip ($N_{x}$) is assumed as infinite so that there is no finite-size effect. Parameters are taken as $m=t=1$, $\lambda=0.2t$, $\Delta_{0}=0.5t$, $\mu=-4.6t$. The angles used are (a) $\theta=0$, (b) $\theta=\theta_{c1}$, (c) $\theta=\theta_{c2}$, (d) $\theta=\frac{\pi}{4}$, (e) $\theta=\frac{\pi}{2}$, (f) $\theta=0$, (g) $\theta=\theta'_{c1}$, (h) $\theta=\theta'_{c2}$, (i) $\theta=\frac{\pi}{4}$, and (j) $\theta=\frac{\pi}{2}$. See main text for the definitions of the angles.}
\end{figure}

While the direct observation of the edge states in terms of ARPES requires very high resolution or large proximity-induced pairing amplitude in the HM film, the STS spectrum can be used as a more sensitive probe to detect the edge states. For clean samples with ideal edges, the STS detects the density of states (DOS) on the edge of the sample, defined as an integration of the spectral function over the edge momenta in the 1D edge BZ \cite{hao16}. As shown in Fig.6 are the results for both $\eta(\mathbf{k})=\cos k_{x}-\cos k_{y}$ and $\eta(\mathbf{k})=\sin k_{x}\sin k_{y}$. Prominent fine structures in the DOS are found, which show striking evolutions with $\theta$ and are related directly to the evolutions in the energy spectra of the strip. There are qualitative differences in the behavior of the DOS between those for $\eta(\mathbf{k})=\cos k_{x}-\cos k_{y}$ and those for $\eta(\mathbf{k})=\sin k_{x}\sin k_{y}$. The behaviors of the two are also very different from the results for the heterostructure with an $s$-wave superconductor \cite{hao16}.

The DOS follows closely the evolution of the edge state spectrum. We emphasize here two most interesting features in the results. The first point is the striking differences in the DOS at $\theta=0$ for the three choices of $\eta(\mathbf{k})$. In contrast to $\eta(\mathbf{k})=\cos k_{x}-\cos k_{y}$ and $\eta(\mathbf{k})=1$ which both have well-defined zero-energy edge states, the zero-energy edge states for $\eta(\mathbf{k})=\sin k_{x}\sin k_{y}$ merge with the bulk Weyl nodes and therefore have vanishing weight on the edge. This explains the finite DOS at $\omega=0$ for $\eta(\mathbf{k})=\cos k_{x}-\cos k_{y}$ and $\eta(\mathbf{k})=1$ \cite{hao16} and the vanishing zero-energy DOS for $\eta(\mathbf{k})=\sin k_{x}\sin k_{y}$. In addition, the edge states for $\eta(\mathbf{k})=\cos k_{x}-\cos k_{y}$ are distorted and have two finite-energy critical points with zero slope [Figs. 4(a) and 4(b)]. They in turn give the two low-energy Van Hove singularities in Fig.6(a). The two low-energy Van Hove singularities in Fig.6(f) are also related to the presence of finite-energy critical points with zero slope in the edge states, as is clear from Figs.5(a) and 5(b). The second feature is the existence of a critical angle at which the edge states on one edge becomes completely (the $s$-wave case) or partly (the two $d$-wave cases) flat. In this aspect, the results for $\eta(\mathbf{k})=\cos k_{x}-\cos k_{y}$ are similar to those for $\eta(\mathbf{k})=1$ \cite{hao16}. The appearance of a nearly flat portion in the edge state dispersion shown in Figs. 4(d) and 4(f) manifest as a prominent zero-energy DOS peak in Figs. 6(b) and 6(c). For $\eta(\mathbf{k})=\sin k_{x}\sin k_{y}$, a portion of the edge state dispersion for one edge also becomes flat at the transition point from the type-I Weyl node to the type-II Weyl node, which would also give a zero-energy Van Hove singularity in the DOS. From Figs. 3(h) and 5(e), $\theta'_{c2}$ is very close to but slightly smaller than this transition point. Correspondingly, we get in Fig.6(h) a not fully developed zero-energy Van Hove singularity. In addition, the Van Hove singularity for $\eta(\mathbf{k})=\sin k_{x}\sin k_{y}$ occurs on the opposite edge as compared to those for $\eta(\mathbf{k})=\cos k_{x}-\cos k_{y}$ and $\eta(\mathbf{k})=1$. This is understood from the fact that, for the edge states close to $k_{y}=0$ on the same edge (e.g., the left edge with $n_{x}=1$) and at $\theta=0$, their velocity for $\eta(\mathbf{k})=\sin k_{x}\sin k_{y}$ is opposite in sign to those for $\eta(\mathbf{k})=\cos k_{x}-\cos k_{y}$ and $\eta(\mathbf{k})=1$.

\subsection{Comparison of energy spectra for three $\eta_{\mathbf{k}}$}

Here, we summarize the main differences between the energy spectra of the heterostructures with $d$-wave form factors [$\eta(\mathbf{k})=\cos k_{x}-\cos k_{y}$ and $\eta(\mathbf{k})=\sin k_{x}\sin k_{y}$] and those for the heterostructure with $s$-wave form factor [$\eta(\mathbf{k})=1$] \cite{hao16}.

For the $s$-wave case, there are three different phases as we change the direction of the magnetization, the fully-gapped phase, the phase without a full gap but has a ubiquitous local gap, and the sWSC phase with a pair of nodes between the two low-energy quasiparticle bands when the magnetization lies along an in-plane direction. For both $\eta(\mathbf{k})=\cos k_{x}-\cos k_{y}$ and $\eta(\mathbf{k})=\sin k_{x}\sin k_{y}$ of the $d$-wave cases, the energy spectra always have nodes and thus describe 2D dWSC.

In terms of the number of Weyl nodes, there is a single pair of Weyl nodes in the sWSC. The two nodes have opposite nonzero quasiparticle energies. For $\eta(\mathbf{k})=\cos k_{x}-\cos k_{y}$, when a nonzero out-of-plane component of the magnetization exists, there are four Weyl nodes. When the magnetization lies completely along an in-plane direction, an additional pair of Weyl nodes emerge and so there are in total three pairs of Weyl nodes. The quasiparticle energies for the nodes are zero for completely out-of-plane magnetization, but are nonzero when an in-plane component of the magnetization exists. For $\eta(\mathbf{k})=\sin k_{x}\sin k_{y}$, there are always two pairs of Weyl nodes in the quasiparticle energy spectrum. The energies of the pair of nodes along $(k_{x},0)$ are fixed at $E=0$ for an arbitrary orientation of the magnetization. For the other pair of nodes along $(0,k_{y})$, the energies are zero for purely out-of-plane magnetization but is nonzero in the presence of a finite in-plane magnetization component.

As regards the type of Weyl nodes, only type-I Weyl nodes are possible for sWSC and dWSC with $\eta(\mathbf{k})=\cos k_{x}-\cos k_{y}$. For $\eta(\mathbf{k})=\sin k_{x}\sin k_{y}$, the pair of nodes fixed at zero energy along $(k_{x},0)$ change between type-I and type-II as the direction of the magnetization varies.

\subsection{Nature of the proximity-induced state in the HM}

In this subsection, we make some comments on the nature of the phases described by Eq.(3), for a HM with spin-singlet pairing term and RSOC. For purely out-of-plane magnetization, the quasiparticle spectrum is similar to a conventional BCS superconductor: A full gap opens for the $s$-wave case which is effectively a spinless $p+ip$ superconductor \cite{hao16}, whereas for the two $d$-wave cases the spectrum consists of four isolated nodes and is gapped everywhere else. As we tune the magnetization away from the out-of-plane direction, a nonzero in-plane component of the magnetization (i.e., $m_{x}\ne0$) cooperates with the RSOC to tilt the quasiparticle spectrum continuously along the $k_{y}$ direction. For the $s$-wave case, along with the transition from the fully-gapped phase to the phase without a full gap but has a ubiquitous local gap, the $E_{1}(\mathbf{k})$ and $E_{-1}(\mathbf{k})$ quasiparticle bands begin to overlap in energy \cite{hao16}. The `Fermi surface' of the quasiparticle bands, constituted by states with quasiparticle energy $E=0$, consists of two disconnected pockets: One `hole'-like pocket from $E_{-1}(\mathbf{k})$ and one `electron'-like pocket from $E_{1}(\mathbf{k})$. For the two $d$-wave cases, while the spectra have nodes for all magnetization orientations, the energy overlap between the $E_{1}(\mathbf{k})$ and $E_{-1}(\mathbf{k})$ quasiparticle bands also increases as the magnetization tilts continuously to the in-plane direction. For $\eta(\mathbf{k})=\cos k_{x}-\cos k_{y}$, the `hole'-like and `electron'-like pockets are always disconnected, similar to the case of $\eta(\mathbf{k})=1$. For $\eta(\mathbf{k})=\sin k_{x}\sin k_{y}$, on the other hand, the `hole'-like and `electron'-like pockets begin to be connected through the two Weyl nodes lying along the $k_{x}$ axis once they turn into type-II.

As shown in Fig.7 is an illustration of the evolution of the $E=0$ Fermi contour explained above, for five typical values of $\theta$ in the range $[0,\frac{\pi}{2}]$. They are interpreted naturally as a series of Lifshitz transitions \cite{lifshitz}. Namely, as the magnetization orientation varies, the $E=0$ Fermi contour undergoes transitions between four and three topologically distinctive configurations for $\eta(\mathbf{k})=\cos k_{x}-\cos k_{y}$ and $\eta(\mathbf{k})=\sin k_{x}\sin k_{y}$, respectively. Similar analysis can be applied to the case of $\eta(\mathbf{k})=1$. It is important to note that the Lifshitz transition is usually encountered and discussed in the normal phase of a metal, as an abrupt change in the Fermi surface topology of the metal \cite{lifshitz,volovik16}. Its revelation in the presence of pairing indicates the essential difference between the state of the present system and other more well-known types of pairings. Also note that, only the Weyl nodes in Figs. 7(i) and 7(j) connect the $n=-1$ and the $n=1$ quasiparticle bands and are type-II, whereas Figs. 7(a)-7(h) contain only type-I Weyl nodes. Therefore, Fig. 7 supplements the definition in Sec. I for the type-I and type-II Weyl nodes.

\begin{figure}[!htb]\label{fig7}
\centering
\includegraphics[width=8.5cm,height=15.0cm]{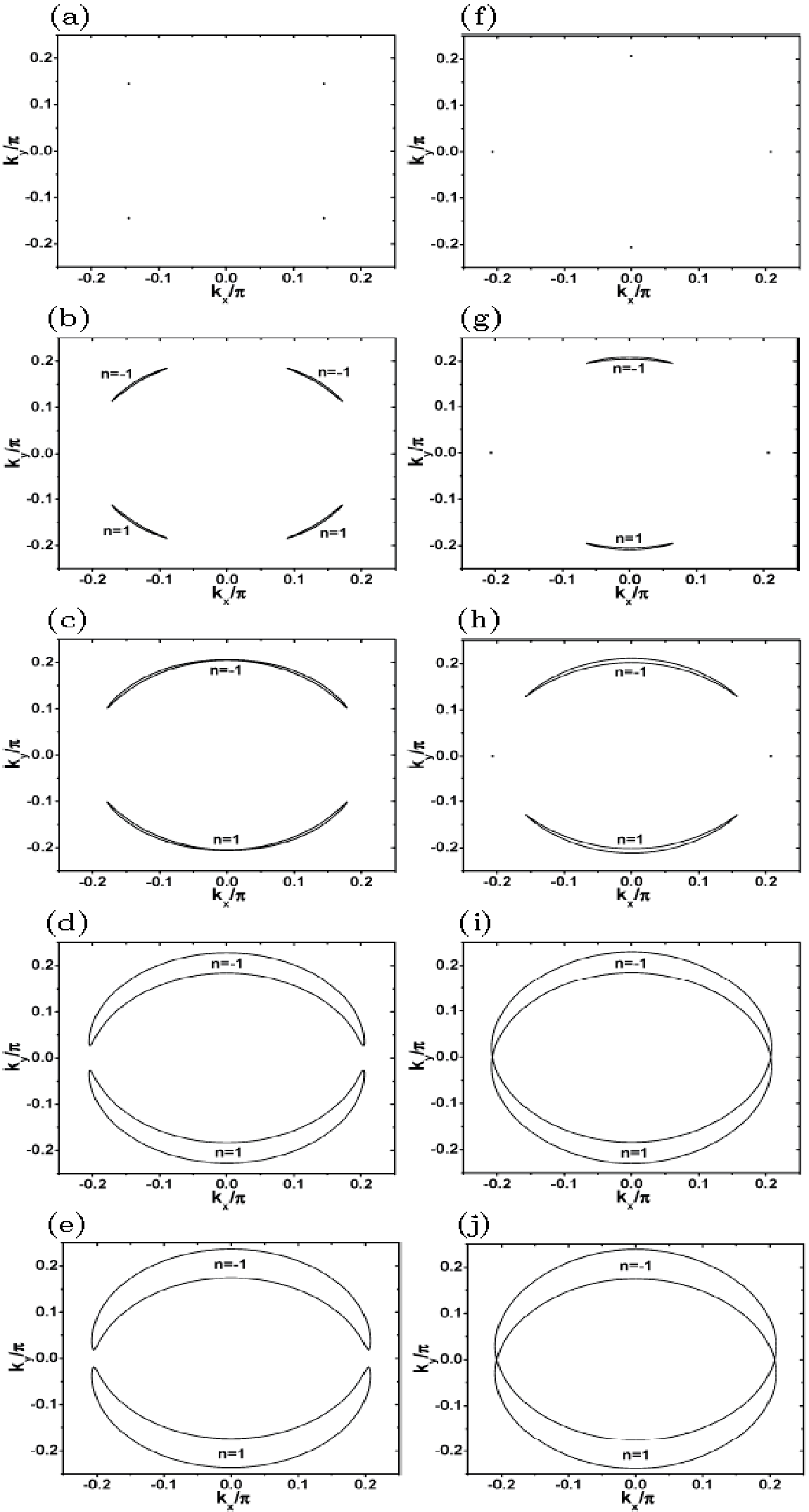}
\caption{The $E=0$ equal-energy contours for the bulk quasiparticle bands. $\eta(\mathbf{k})=\cos k_{x}-\cos k_{y}$ for (a, b, c, d, e) and $\eta(\mathbf{k})=\sin k_{x}\sin k_{y}$ for (f, g, h, i, j). Parameters are taken as $m=t=1$, $\lambda=0.2t$, $\Delta_{0}=0.5t$, $\mu=-4.6t$. The angles used are (a) $\theta=0$, (b) $\theta=\theta_{c1}$, (c) $\theta=\theta_{c2}$, (d) $\theta=\frac{\pi}{4}$, (e) $\theta=\frac{\pi}{2}$, (f) $\theta=0$, (g) $\theta=\theta'_{c1}$, (h) $\theta=\theta'_{c2}$, (i) $\theta=\frac{\pi}{4}$, and (j) $\theta=\frac{\pi}{2}$. See main text for the definitions of the angles. The numberings of the quasiparticle bands to which the pockets belong are indicated in the figures.}
\end{figure}

Now, we compare the quasiparticle spectrum for the present state and those for several well-known types of pairings, including the BCS-like pairing \cite{bcs}, the FFLO pairing \cite{ff,lo}, and the so-called breached pairing (or, Sarma state) \cite{liu03,gubankova03,bedaque03,sarma63}. The BCS pairing naturally occurs in a system with a centro-symmetric Fermi surface. The pairing between states of equal and opposite momenta thus gives a spectrum in which the $E_{1}(\mathbf{k})$ quasiparticle band is completely empty and the $E_{-1}(\mathbf{k})$ quasiparticle band is fully occupied. For our case, the pairing is also between states of equal and opposite momenta. But the energy spectrum deviates significantly from the conventional BCS spectrum that is symmetric with respect to $E=0$ for each $\mathbf{k}$. In the FFLO pairing, the net momentum of a pair is nonzero. A simplest case where this phase might be found is when the Fermi surface is asymmetric with respect to the center of the BZ. In our case, while the Fermi surface is asymmetric in the presence of an in-plane component of the exchange field, the net momentum of a Cooper pair is always zero. Finally, in the breached pairing state, the pairing gap between \emph{two} bands of unequal Fermi momenta opens at nonzero energy. For our system, in the presence of an in-plane component of the exchange field and away from $k_{y}=0$, the pairing gap between the $E_{1}(\mathbf{k})$ and $E_{-1}(\mathbf{k})$ bands also opens at nonzero energy and is similar to the breached pairing state. However, on one hand the present system consists of only one Fermi surface in the normal phase, on the other hand the states on the Fermi surface with $k_{y}\simeq0$ are usually gapped at $E=0$. Only for $\eta(\mathbf{k})=\sin k_{x}\sin k_{y}$ and when the two nodes along $(k_{x},0)$ have turned from type-I to type-II, would the full Fermi surface be gapless and the spectrum be most similar to the breached pairing state. However, the $E=0$ states in this case consist of both `electron'-like states from $E_{1}(\mathbf{k})$ and `hole'-like states from $E_{-1}(\mathbf{k})$, rather than from a single quasiparticle band in the case of a standard breached pairing state.

From the above comparisons, we see that the appearance of the unique pairing state in the present system is caused by a synergic action of two factors: The BCS-like pairing correlation induced through proximity effect, and the asymmetry of the Fermi surface in the HM. The first aspect of the model is a result of conservation of the parallel component of the momentum through the interface which is assumed to be perfectly translational invariant. The second factor results from a combination of the in-plane component of the magnetization and the RSOC.

In our treatment of the pairing in the HM, we have assumed that it arises completely from the proximity coupling with the spin-singlet superconductor. If instead we assume that there is an intrinsic pairing tendency in the HM, then a self-consistent calculation is necessary in addition to the proximity effect. In this case, in the presence of an in-plane component of the magnetization which drives the Fermi surface asymmetric with respect to the BZ center, the pairing term obtained in a self-consistent manner could also possibly be the FFLO pairing \cite{buzdin90,radovic91,demler97,eschrig03,buzdin05}.

\section{topological properties}

In Ref.\cite{hao16} focusing on the heterostructures with $s$-wave superconductor, we have characterized the phases and phase transitions driven by $\theta$ in terms of four topological quantities: A Pfaffian $Z_{2}$ invariant defined based on the particle-hole symmetry of the model, the TKNN number defined in terms of the occupied quasiparticle states, the band-wise Chern numbers associated with separate nondegenerate quasiparticle bands, and the Zak phase which is a Berry phase defined along one-dimensional subspace of the 2D BZ \cite{hao16}. The band-wise Chern numbers in the $s$-wave case are motivated from the ubiquitous local gaps in the quasiparticle spectrum, except for $\theta=\frac{\pi}{2}$ and $\theta=\frac{3\pi}{2}$ which correspond to the 2D Weyl superconductivity phase \cite{hao16}. In the case of heterostructure with $d$-wave superconductor, however, there are always nodes between the two low-energy quasiparticle bands $E_{-1}(\mathbf{k})$ and $E_{1}(\mathbf{k})$. Therefore, the band-wise Chern number cannot be defined for the $d$WSCs. In addition, the zero-energy nodes existent for out-of-plane magnetization for $\eta(\mathbf{k})=\cos k_{x}-\cos k_{y}$ and always existent for  $\eta(\mathbf{k})=\sin k_{x}\sin k_{y}$ also cause difficulty in calculating the TKNN number. Thus, we focus on the Pfaffian $Z_{2}$ invariant and the Zak phase to characterize the topological properties of the phases in $d$WSCs.

Before studying the topological invariants, we list the relevant symmetries (or, broken symmetries) of the model defined by Eq.(3). One important aspect of the model is the absence of time-reversal symmetry, $\Theta=-i\tau_{0}\sigma_{2}K$, where $K$ denotes complex conjugation. Once $m\ne0$, no matter what the value of $\theta$ is, we have $\Theta^{-1}h(\mathbf{k},m_{x},m_{z})\Theta=h(-\mathbf{k},-m_{x},-m_{z})$ and thus the time-reversal symmetry is broken. Besides, the inversion symmetry, which for the present single-orbital model is simply the operation bringing $\mathbf{k}$ to $-\mathbf{k}$, is broken by the RSOC since we have $h(-\mathbf{k},\lambda)=h(\mathbf{k},-\lambda)\ne h(\mathbf{k},\lambda)$. However, the model defined by Eq.(3) always has the particle-hole symmetry, $\Xi=\tau_{1}\sigma_{0}K$. That is,
\begin{equation}
\Xi^{-1}h(\mathbf{k})\Xi=-h(-\mathbf{k})
\end{equation}
holds for all three $\eta(\mathbf{k})$ and for all values of $\theta$. The next relevant symmetry is the mirror reflection symmetry. For the mirror reflection with respect to the plane perpendicular to the $x$ axis, the corresponding mirror reflection operator is $\tilde{M}_{x}=i\tau_{3}\sigma_{1}$. Similarly, we have $\tilde{M}_{y}=i\tau_{0}\sigma_{2}$ and $\tilde{M}_{z}=i\tau_{3}\sigma_{3}$, for mirror reflections with respect to the planes perpendicular to the $y$ axis and $z$ axis. For $\eta(\mathbf{k})=\cos k_{x}-\cos k_{y}$ and $\eta(\mathbf{k})=1$, we have
\begin{equation}
\tilde{M}^{-1}_{x}h(\mathbf{k})\tilde{M}_{x}=h(-k_{x},k_{y},-m_{z}),
\end{equation}
\begin{equation}
\tilde{M}^{-1}_{y}h(\mathbf{k})\tilde{M}_{y}=h(k_{x},-k_{y},-m_{x},-m_{z}),
\end{equation}
and
\begin{equation}
\tilde{M}^{-1}_{z}h(\mathbf{k})\tilde{M}_{z}=h(-k_{x},-k_{y},-m_{x}).
\end{equation}
For $\eta(\mathbf{k})=\sin k_{x}\sin k_{y}$, by redefining the mirror reflection operator for $x$ and $y$ directions as $\tilde{M}'_{x}=i\tau_{0}\sigma_{1}$ and $\tilde{M}'_{y}=i\tau_{3}\sigma_{2}$, it is easy to see that the BdG Hamiltonian $h(\mathbf{k})$ satisfies the same transformation properties as the other two pairing channels. From Eqs.(16)-(18), the model has the mirror symmetry along $x$ only when $m_{z}=0$. No mirror symmetry along $y$ exists, once $m\ne0$. When $m_{x}=0$, $\tilde{M}_{z}$ acts effectively as an inversion symmetry of the model. This effective inversion symmetry ensures that each pair of Weyl nodes having opposite wave vectors can share the same quasiparticle energy for $\theta=0$ or $\pi$. Finally, the model has the four-fold rotation symmetry around the $z$ axis when $m_{x}=0$, namely
\begin{equation}
\tilde{S}^{-1}_{z}h[R_{\frac{\pi}{2}}(\mathbf{k})]\tilde{S}_{z}=h(\mathbf{k}).
\end{equation}
$R_{\frac{\pi}{2}}(\mathbf{k})=(k_{y},-k_{x})$. $\tilde{S}_{z}=e^{i\frac{\pi}{2}\frac{\tau_{3}\sigma_{3}}{2}}=\frac{1}{\sqrt{2}}(\tau_{0}\sigma_{0}+i\tau_{3}\sigma_{3})$ for $\eta(\mathbf{k})=1$. $\tilde{S}_{z}=\tau_{3}e^{i\frac{\pi}{2}\frac{\tau_{3}\sigma_{3}}{2}}=\frac{1}{\sqrt{2}}(\tau_{3}\sigma_{0}+i\tau_{0}\sigma_{3})$ for $\eta(\mathbf{k})=\cos k_{x}-\cos k_{y}$ and $\eta(\mathbf{k})=\sin k_{x}\sin k_{y}$. As a result of this four-fold rotation symmetry, all four Weyl nodes share the same quasiparticle spectrum for $\theta=0$ and $\pi$. Other symmetries (e.g., the so-called magnetic mirror reflection symmetries \cite{lu15}) of the model, while on one hand can be defined as a combination of several of the above symmetries, on the other hand are not directly relevant to the following discussions. Therefore, we do not list them here.

\subsection{Pfaffian $Z_2$ invariant}

From the particle-hole symmetry of the model defined by Eq.(3), the Pfaffian $Z_2$ invariant for the dWSCs can be defined as \cite{ghosh10,hao16}
\begin{equation}
P=\text{sgn}[|\Delta(\mathbf{k}=(0,0))|^{2}+\epsilon^{2}_{\mathbf{k}=(0,0)}-m^{2}],
\end{equation}
where $\epsilon_{\mathbf{k}=(0,0)}=-4t-\mu$, the function $\text{sgn}(x)$ gives the sign of a real number $x$. For both $\eta(\mathbf{k})=\cos k_{x}-\cos k_{y}$ and $\eta(\mathbf{k})=\sin k_{x}\sin k_{y}$, we have $\Delta(\mathbf{k}=(0,0))=0$. For the present parameters with large $m$ and a single nondegenerate Fermi surface, we have $m>|\mu+4t|$, the phase is therefore always nontrivial ($P=-1$) from the Pfaffian $Z_{2}$ invariant.

If the energy spectrum of the particle-hole symmetric BdG Hamiltonian, for example Eq.(3) with $\eta(\mathbf{k})=1$ and $\theta=0$, is fully gapped, the Pfaffian $Z_2$ invariant was shown by Ghosh \emph{et al} to give the parity of the Chern number \cite{ghosh10}. In another work based on a model which can be considered as the $\theta=0$ case of Eq.(3) with $\eta(\mathbf{k})=\cos k_{x}-\cos k_{y}$ or $\eta(\mathbf{k})=\sin k_{x}\sin k_{y}$, Sato and Fujimoto pointed out that although the spectrum is gapless, the parity of the Chern number is still well-defined and determined by the same formula [i.e., Eq.(20)] for the Pfaffian $Z_2$ invariant \cite{sato10,sato10b}. Therefore, for both of the two choices of $\eta(\mathbf{k})$ for the $d$WSCs and for purely out-of-plane magnetization, the parity of the Chern number is well-defined and is odd for the parameters considered in Sec.III. From the topological nature of $P$, it is then natural to expect that the system described by Eq.(3) retains a certain topological character as $\theta$ deviates from $0$ and $\pi$. This is confirmed from the persistence of the edge states shown in Sec.III, except for Fig.3(j) where the edge modes are merged to the bulk bands. From the analysis of the Zak phase in Sec.IV C, however, there should be hidden edge states for the parameters of Fig.3(j).

\subsection{Chiral edge states}

The edge states for a general $\theta$ evolve continuously from those for $\theta=0$. So, it is important to define the chirality of the edge states for $\theta=0$. In the simplest case of a $p+ip$ superconductor, the edge states are linearly dispersive and counter-propagating on the two opposing edges. The chirality of the edge state can be defined in terms of the sign of the velocity of the edge modes. The HM layer of the heterostructure with $s$-wave superconductor is equivalent to a $p+ip$ superconductor, the chirality of the edge states can thus be defined in this manner \cite{hao16}. In the heterostructures with $d$-wave superconductors, however, the edge states is curvous and the velocity changes sign several times (see Figs. 2 and 3). Nonetheless, we can still define the chirality of the edge states in terms of the velocity of the edge modes in the following manner. On an edge with a unique edge state, we divide the full dispersion of it into several segments connected together through points with zero velocity. Each segment is assigned a chirality, positive or negative, equal to the sign of the velocity in this section of the dispersion. The overall chirality is defined simply as a multiplication of the chiralities of all the segments. In this manner, it is straightforward to see from Figs. 4 and 5 in Sec. III C (and, Fig. 5 in Ref.\cite{hao16} for the $s$-wave case) that the edge states on the same edge and for $\theta=0$ have the same chirality for all three $\eta(\mathbf{k})$. In addition, as $\theta$ increases, the chirality of the edge states on the left edge ($n_{x}=1$) keeps unchanged, whereas the chirality of the edge states on the right edge ($n_{x}=N_{x}$) change sign. The $\theta$-dependence of the (generalized) chirality of the edge states for all three pairings are thus qualitatively the same.

\subsection{Zak phase}

The Zak phase is a Berry phase defined as a line integral over the Berry connection along a 1D loop of the 2D BZ \cite{zak89,delplace11}. In calculation, one component (i.e., $k_{x}$ or $k_{y}$) of the wave vector is fixed as a parameter, integration over the wave vector along the orthogonal direction then traces a 1D loop, as a result of the periodicity of the BZ. For example, by taking $k_{y}$ as a parameter, the Zak phase for the $n$-th ($n=-2$, $-1$, $1$, $2$) quasiparticle band is defined as
\begin{equation}
Z_{n}(k_{y})=\int_{-\pi}^{\pi} dk_{x}A^{(x)}_{n}(k_{x},k_{y}),
\end{equation}
where $A^{(x)}_{n}(k_{x},k_{y})$ is the $x$ component of the Berry connection for the $n$-th quasiparticle band, $\mathbf{A}_{n}(\mathbf{k})=i<u_{n}(\mathbf{k})|\boldsymbol{\nabla}_{\mathbf{k}}u_{n}(\mathbf{k})>$. $|u_{n}(\mathbf{k})>$ is the
eigenvector of the $n$-th quasiparticle band. Another set of Zak phase, $Z_{n}(k_{x})$, is defined in the same manner by reversing the roles of $k_{x}$ and $k_{y}$. When the effective 1D model is inversion symmetric, the Zak phase can only take two values ($0$ or $\pi$), otherwise the Zak phase can take any value between $0$ and $2\pi$ \cite{zak89,delplace11}. For the $s$WSC defined by $\eta(\mathbf{k})=1$ and $m_{z}=0$, the relevant symmetry is $\tilde{M}^{-1}_{x}h(\mathbf{k})\tilde{M}_{x}=h(-k_{x},k_{y})$. The positions of the two Weyl nodes of the $s$WSC can be captured by a jump in the Zak phase between $0$ and $\pi$ \cite{hao16}. In addition, on an edge running along $y$ (so that $k_{y}$ is still a good quantum number but $k_{x}$ is not), there are edge states in the segment of $k_{y}$ for which the Zak phase is $\pi$ \cite{hao16}.

From Eqs.(16)-(18), the $d$WSCs with $\eta(\mathbf{k})=\cos k_{x}-\cos k_{y}$ and $\eta(\mathbf{k})=\sin k_{x}\sin k_{y}$ have the same mirror symmetry (or, modified mirror symmetry, for $\eta(\mathbf{k})=\sin k_{x}\sin k_{y}$) as $\eta(\mathbf{k})=1$. As a result, the cases in which the Zak phase can take a well-defined quantized value of $0$ or $\pi$ are still associated with $\theta=\frac{\pi}{2}$ or $\frac{3\pi}{2}$, when the magnetization has only an in-plane component. In this case, the quantization of $Z_{n}(k_{y})$ follows from Eq.(16). In addition, by combining the particle-hole symmetry $\Xi$ and $\tilde{M}_{x}$, we can see that the model has a `inversion-like' symmetry along $k_{y}$ for fixed $k_{x}$. More exactly, we have
\begin{equation}
\Xi^{-1}\tilde{M}^{-1}_{x}ih(\mathbf{k})\tilde{M}_{x}\Xi=ih(k_{x},-k_{y},-m_{z}).
\end{equation}
Therefore, the Zak phase defined along $k_{y}$ loops should also take quantized values ($0$ or $\pi$). On the other hand, once the out-of-plane component of the magnetization ($m_{z}$) is nonzero, no `inversion-like' symmetry can be defined along an arbitrary 1D loop with one wave vector fixed at a general value. A few exceptions to this conclusion are for $m_{x}=0$ and for those loops with the fixed wave vector taking time-reversal-invariant values (i.e., $0$ and $\pm\pi$). For these special loops, the 1D effective model has an `inversion-like' symmetry enforced by the symmetry of $\tilde{M}_{z}$ defined by Eq.(18).

\begin{figure}[!htb]\label{fig8}
\centering
\includegraphics[width=6.5cm,height=8.5cm]{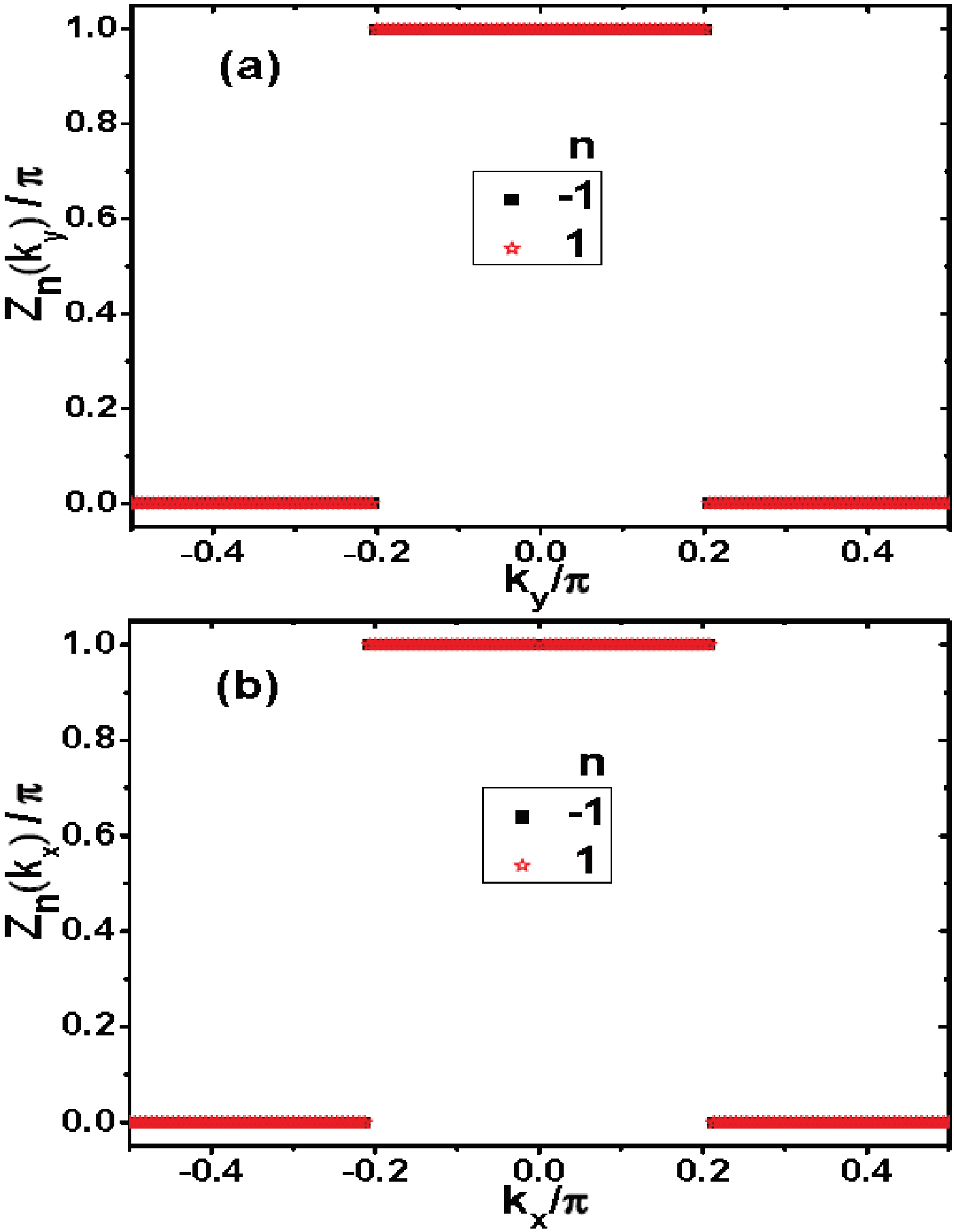}
\caption{(Color online) Nonzero Zak phases for the two low-energy quasiparticle bands $E_{\pm1}(\mathbf{k})$, at $\theta=\pi/2$. $\eta(\mathbf{k})=\cos k_{x}-\cos k_{y}$ for (a), and $\eta(\mathbf{k})=\sin k_{x}\sin k_{y}$ for (b). The parameters are the same as those used in Sec.III.}
\end{figure}

We have calculated explicitly the Zak phases $Z_{n}(k_{x})$ and $Z_{n}(k_{y})$ for both $\eta(\mathbf{k})=\cos k_{x}-\cos k_{y}$ and $\eta(\mathbf{k})=\sin k_{x}\sin k_{y}$, at $\theta=0$ and $\theta=\pi/2$. In agreement with the above expectation based on symmetry, $Z_{n}(k_{x})$ and $Z_{n}(k_{y})$ take quantized values of $0$ or $\pi$ for $\theta=\pi/2$. For $\theta=0$, the Zak phases are quantized only when the fixed wave vector takes $0$ or $\pm\pi$. We show in Figure $8$ the nontrivial results for the Zak phase, which are $Z_{n}(k_{y})$ for $\eta(\mathbf{k})=\cos k_{x}-\cos k_{y}$ [Fig.8(a)] and $Z_{n}(k_{x})$ for $\eta(\mathbf{k})=\sin k_{x}\sin k_{y}$ [Fig.8(b)], both at $\theta=\pi/2$. Only the Zak phases for the two low-energy quasiparticle bands (namely, $n=\pm1$) are shown since the results for the two high energy bands are always zero. Fig.8(b) does not contain results for $k_{x}=0$ since the integration loop for $k_{x}=0$ contain two nodes between $E_{-1}(\mathbf{k})$ and $E_{1}(\mathbf{k})$ which makes the Zak phase ill defined. In Fig.8(a), results for the loops containing the nodes are also excluded. In addition, we have obtained (at $\theta=\pi/2$) $Z_{n}(k_{x})=0$ for $\eta(\mathbf{k})=\cos k_{x}-\cos k_{y}$ and $Z_{n}(k_{y})=0$ for $\eta(\mathbf{k})=\sin k_{x}\sin k_{y}$.

Comparing Fig.8(a) with the Zak phase calculated for the $s$WSC \cite{hao16}, it is clear that they are essentially the same. Although the pairing amplitudes are taken at different values, the positions of the Weyl nodes of the $s$WSC and the new pair of Weyl nodes for the $d$WSC with $\eta(\mathbf{k})=\cos k_{x}-\cos k_{y}$ are identical to each other, because they are determined by the vanishing of the effective pairing amplitude. And, though there are two additional pairs of Weyl nodes from $\eta(\mathbf{k})=0$, the Zak phase do not see them. This is also a very natural result, though, because each node contributes $\pi$ or $0$ to the Zak phase, so the net change of the Zak phase upon crossing the line with two Weyl nodes is zero (modulo $2\pi$). In addition, it is clear that the region with a nontrivial Zak phase of $\pi$ is also characterized by the presence of edge modes connecting the two additional pairs of Weyl nodes emerging from the emergent mirror symmetry of the model. On the other hand, the fact that $Z_{n}(k_{x})=0$ for all $k_{x}$ is consistent with the fact that the change in the number of Weyl nodes over which the integration loop sweeps is always an even number, zero or two. Therefore, even if each node contributes $\pi$, the total change in the Zak phase defined modulo $2\pi$ is still zero.

The Zak phase for $\eta(\mathbf{k})=\sin k_{x}\sin k_{y}$ is however more peculiar. Namely, $Z_{n}(k_{y})=0$ although the two Weyl nodes along $(0,k_{y})$ are separated from each other and thus the integration loop sweeps only a single Weyl node. This result can be understood from the low-energy effective models for the two nodes along $(0,k_{y})$ for $\eta(\mathbf{k})=\sin k_{x}\sin k_{y}$: They are real symmetric [see Eqs. (A11)-(A13) in Appendix A] and thus the wave function can be taken as purely real. As such, no nontrivial phase factor can be collected when we traverse the nodes. On the other hand, the low-energy effective models for the two Weyl nodes of $\eta(\mathbf{k})=\sin k_{x}\sin k_{y}$ along $(k_{x},0)$ (which are type-II for $\theta=\pi/2$) are complex. This explains why the Zak phase $Z_{n}(k_{x})$ shows a jump of $\pi$ when the integration path traverse one node. $Z_{n}(k_{y})=0$ is also in agreement with the absence of well-defined edge states in Fig.3(j) for edges running along the $y$ direction. On the other hand, from the nontrivial wave vector dependence of $Z_{n}(k_{x})$, it is natural to expect the presence of edge states on the edges running along $x$-direction. This in turn can account for the nontrivial result of the Pfaffian $Z_{2}$ invariant. However, since the bulk quasiparticle spectrum is completely gapless from the $k_{x}$ direction, the edge states, even if they exist, would merge to the bulk spectra and become undiscernible. However, it is still possible that the nontrivial Zak phase $Z_{n}(k_{x})$ can have some influence on the properties of the system. We leave the analysis over this interesting possibility to later works.

Finally, we consider the case of $\theta=0$. Consistent with the symmetry analysis, we have found that the Zak phase $Z_{n}(k_{y})$ [$Z_{n}(k_{x})$] for $\theta=0$ does not take quantized value for a general $k_{y}$ ($k_{x}$). On the other hand, the Zak phase corresponding to wave vectors $0$ and $\pm\pi$ do take quantized values ($0$ or $\pi$). These quantized Zak phase can be utilized to define a new  $Z_{2}$ invariant,
\begin{equation}
P'_{\alpha}=e^{iZ_{1}(k_{\alpha}=0)}e^{iZ_{1}(k_{\alpha}=\pi)},
\end{equation}
where $\alpha=x$ or $y$ when considering an edge running along $\alpha$ direction. $P'_{\alpha}=-1$ and $1$ corresponds to the existence and nonexistence of well-defined edge state at $k_{\alpha}=0$. This connection can be verified easily by making a comparison between the energy spectra shown in Sec.III and the following numerical results for the Zak phase, at $\theta=0$: For $\eta(\mathbf{k})=\cos k_{x}-\cos k_{y}$, $Z_{1}(k_{y}=\pi)=Z_{1}(k_{x}=\pi)=0$ and $Z_{1}(k_{y}=0)=Z_{1}(k_{x}=0)=\pi$. For $\eta(\mathbf{k})=\sin k_{x}\sin k_{y}$, $Z_{1}(k_{y}=\pi)=Z_{1}(k_{x}=\pi)=Z_{1}(k_{y}=0)=Z_{1}(k_{x}=0)=0$. For $\eta(\mathbf{k})=\sin k_{x}\sin k_{y}$, this result is consistent with the fact that since for both edges running along $x$ and $y$, there are always projections of two Weyl nodes to the origin of the 1D BZ, which act as bulk channels coupling the two edges together, no well-defined edge states can be defined at the center of the one dimensional edge BZ.

In previous works, the edge states of a two-dimensional superconductor have also been characterized by a one-dimensional winding number \cite{sato11}. The winding number is defined only in the presence of the chiral symmetry. Our model breaks explicitly the time-reversal symmetry and therefore does not have chiral symmetry. However, for the special cases (purely in-plane or purely out-of-plane magnetization) where the Zak phase can have quantized values ($0$ or $\pi$), we can combine mirror reflection and time-reversal symmery and define the so-called magnetic mirror reflection symmetry \cite{lu15}. This, in combination with the particle-hole symmetry of the model, gives us a generalized chiral symmetry which can be used to define the corresponding winding number. In this regard, the Zak phase and the one-dimensional winding number should give equivalent topological characterizations to the system.

\section{effect of partial spin polarization}

In the above discussions, we have focused on heterostructures in which the HM has a single nondegenerate Fermi surface. It is interesting to ask what if the spin polarization is not complete, namely what if the two spin-split bands both contribute to the Fermi surface \cite{zhu00}. This is relevant since ferromagnetic materials with partially polarized Fermi surface is more commonly found than HM. This problem is also relevant if we turn our attention from the HM layer to the layer of spin-singlet superconductor. It is clear that the same model can be used to describe this layer. The only qualitative difference is just in the magnitude of the exchange field, since the proximity-induced exchange field in the $d$-wave superconductor is in general not large enough to split completely one spin component off the Fermi surface.

Consider a ferromagnetic metal with a Fermi surface consisting of two sheets with opposite spin-polarization and different volumes. The two sheets with opposite spin polarizations would naturally be separated from each other. Replacing the HM in the heterostructure studied above with such a partially polarized ferromagnet, we can expect to apply the previous analyses separately to each spin component. As such, we can expect to get two sets of low-energy effective models with similar properties but related separately to the two spin-split bands. In particular, for each band, there will be a pair of counter-propagating chiral Majorana edge states. It is then interesting to ask whether or not the two sets of chiral edge modes are always separated from each other. Or, in cases where they coexist in some parameter ranges, how would they interact with each other.

\begin{figure}[!htb]\label{fig9}
\centering
\includegraphics[width=6.5cm,height=12.5cm]{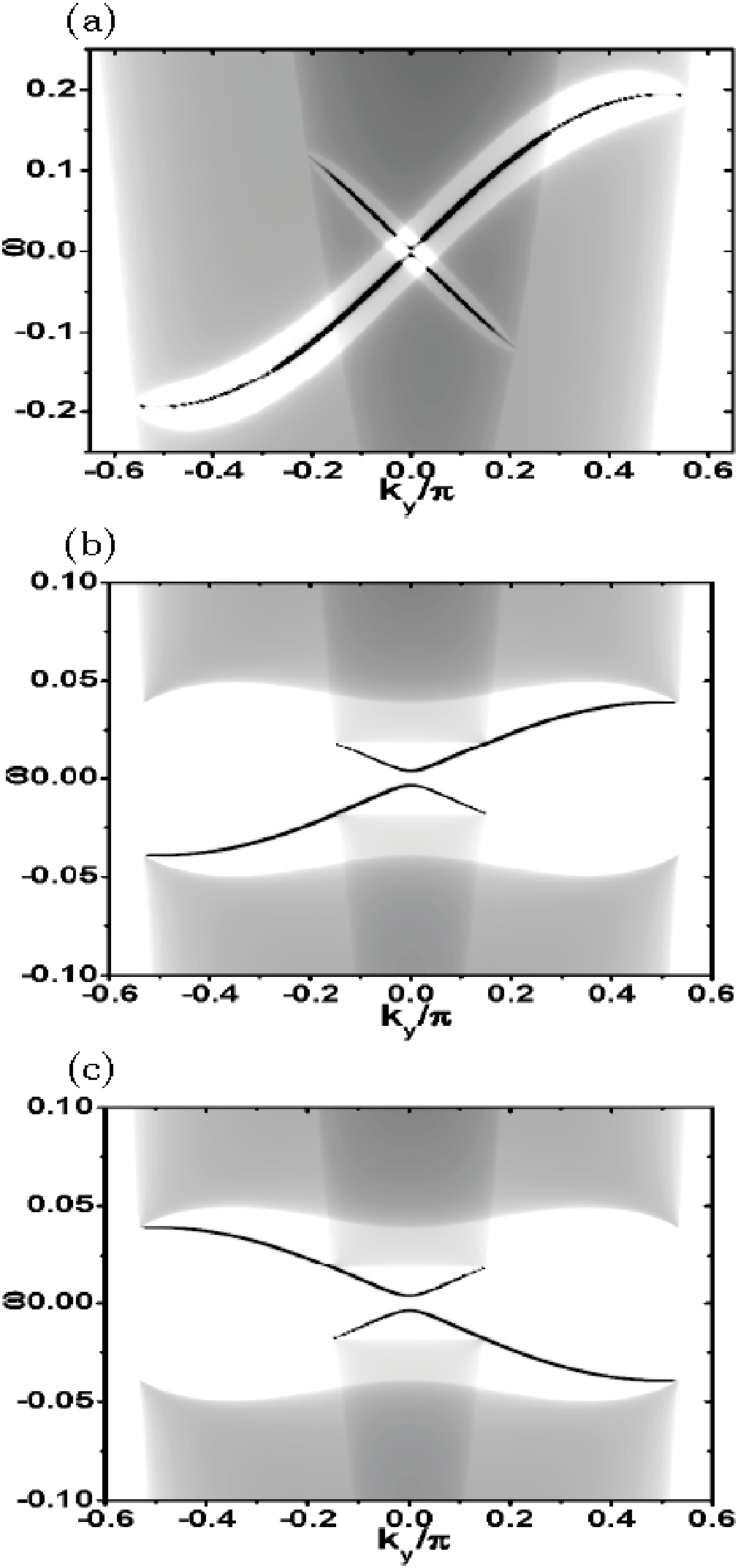}
\caption{Spectral functions for the two edges of a strip of the system described by Eq.(3), with $\eta(\mathbf{k})=1$. (a) $\theta=\pi/2$ which corresponds to an $s$WSC, results for the two edges are the same. (b) and (c) are spectral functions on the left ($n_{x}=1$) and right ($n_{x}=N_{x}$) edges of the strip for $\theta=0$. The parameters used are $t=1$, $m=t$, $\lambda=0.2t$, $\Delta_{0}=0.2t$, $\mu=-2.6t$. The darker the color, the larger the spectral weight.}
\end{figure}

To answer the above question, we have studied the changes in the properties of the Weyl superconductivity phase brought by increasing the chemical potential so that it crosses both of the two spin-polarized subbands of the ferromagnet. As shown in Fig.9(a) are the spectral functions for the edge layers (results for the two edges are the same) of a strip for $\eta(\mathbf{k})=1$ and $\theta=\pi/2$, namely the $s$WSC. While on one hand the edge modes for the two spin species keep largely independent of each other, a hybridization gap opens at ($k_{y}=0$, $\omega=0$). Besides, because the Fermi sheets of the two spin channels open gaps independently, the edge modes in the non-overlapping regions of the pairing gaps of the two pin-split bands will be masked by the bulk states of the opposite-spin band. Overall, the chiral edge modes appear in \emph{three disconnected} regions of the local energy gap of the energy spectrum. In Fig.10, we show the spectral functions on the two edges of a strip of the $d$WSC at $\theta=0$ for $\eta(\mathbf{k})=\cos k_{x}-\cos k_{y}$ [10(a) and 10(b)] and $\eta(\mathbf{k})=\sin k_{x}\sin k_{y}$ [10(c) and 10(d)]. The results for $\eta(\mathbf{k})=1$ at $\theta=0$ are shown as Figs.9(b) and 9(c). For $\eta(\mathbf{k})=\cos k_{x}-\cos k_{y}$, we observe similar coexistence of two chiral edge states and the opening of a hybridization gap at ($k_{y}=0$, $\omega=0$) between the two edge states belonging separately to the two spin-subbands. For $\eta(\mathbf{k})=\sin k_{x}\sin k_{y}$, the two copies of edge states simply superimpose together and no hybridization gap opens. In Ref.\cite{sato10}, it was pointed out that the topological phase with a single Majorana edge state can be obtained for vanishingly small exchange field when the chemical potential is at the band bottom ($\mu=-4t$). While that case still has a single band crossing the chemical potential, our present case has two bands contributing to the Fermi surface. From the numerical results in Fig.10 and comparing it with Fig.9 for the $s$-wave case, we see that while the edge states exist when the Fermi surface consists of two sheets, there is a pair of counter-propagating edge modes on each edge which usually hybridize and open a gap. The physical picture for the $d$-wave case is therefore of no essential difference from the $s$-wave case as regards the number of edge states and their hybridization. Since there are even number of edge states on each edge, they are not topologically protected. This is consistent with the Pfaffian $Z_{2}$ invariant defined in Eq.(20), which gives the topologically trivial result of $P=1$ for the present set of parameters. The phase on the superconductor side of the heterostructure is therefore topologically trivial. Topologically nontrivial physics are located in the HM layer.

\begin{figure}[!htb]\label{fig10} \centering
\includegraphics[width=8.5cm,height=7.0cm]{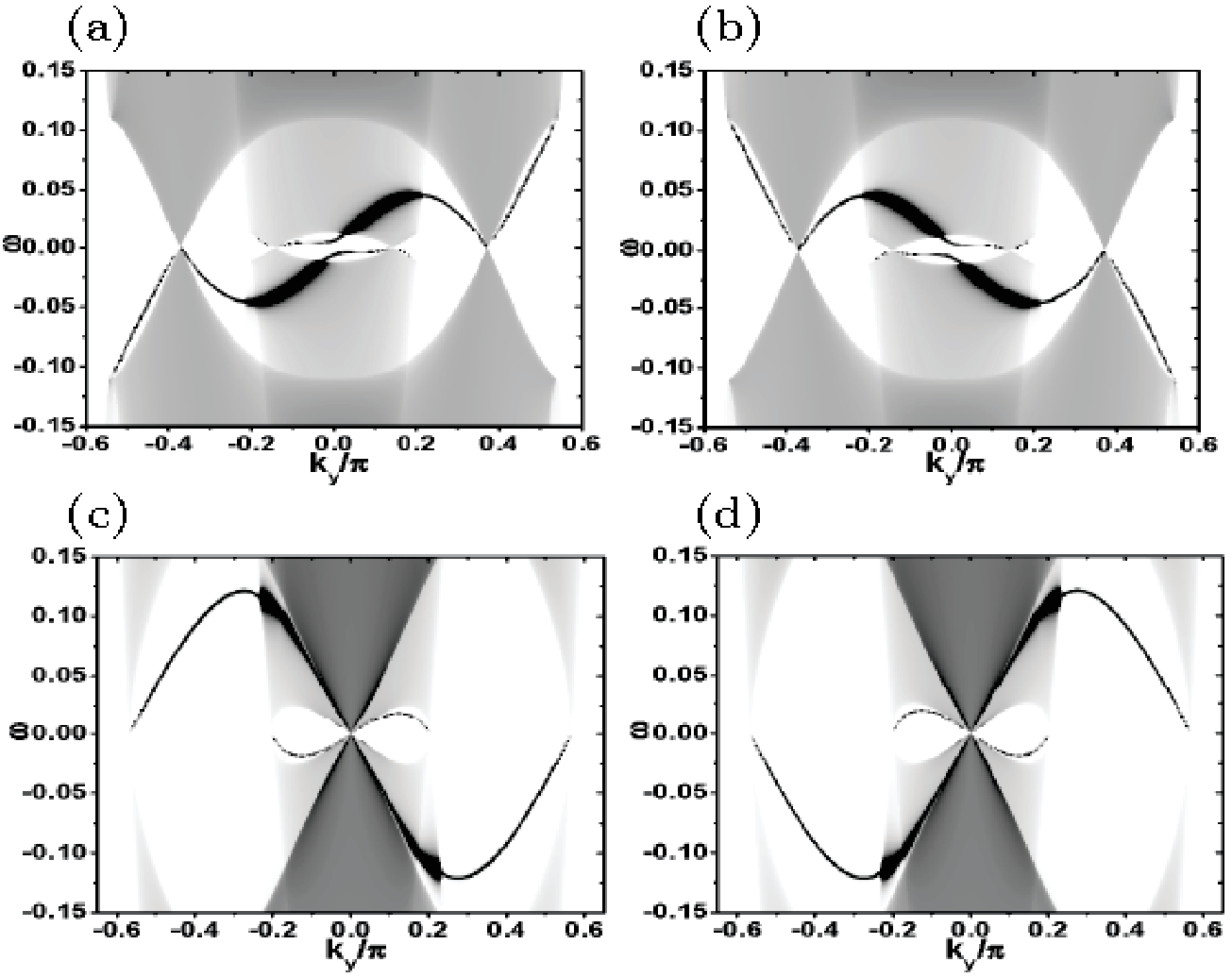}
\caption{Spectral functions for the left (a, c) and right (b, d) edges of a strip of the heterostructure described by Eq.(3), with $\eta(\mathbf{k})=\cos k_{x}-\cos k_{y}$ (a, b) and $\eta(\mathbf{k})=\sin k_{x}\sin k_{y}$ (c, d). The angle is taken as $\theta=0$. The parameters are chosen as $t=1$, $m=t$, $\lambda=0.2t$, $\Delta_{0}=0.5t$, $\mu=-2.6t$. Darker color means larger spectral function.}
\end{figure}

\section{summary}

In summary, we have predicted the realization of two-dimensional Weyl superconductivity phases in a heterostructure consisting of a half-metal (or a highly spin-polarized ferromagnet) clamped in between an insulating substrate and a spin-singlet superconductor. When the superconductor has $d$-wave symmetry (with a symmetry factor of $d_{x^{2}-y^{2}}$ or $d_{xy}$), the two-dimensional Weyl superconductivity phase is realized for an arbitrary magnetization direction in the half-metal. Quasiparticle energy spectra are studied which exhibit interesting variations with the direction of the magnetization. In particular, we have predicted the transition between type-I and type-II Weyl nodes by changing the magnetization orientation. In addition, the evolution of the phase with the magnetization orientation is unveiled through the Lifshitz transitions in the zero-energy quasiparticle spectrum. The phases are shown to be qualitatively different from several well-known types of pairings. We characterize the topology of phases by the Pfaffian $Z_{2}$ invariant and the Zak phase.

\begin{acknowledgments}
We thank Jun Wang and Zhi-Ping Niu for helpful discussions. This work was supported by the Texas Center for Superconductivity at the University of Houston and the Robert A. Welch Foundation (Grant No. E-1146). L.H. is also supported by the National Natural Science Foundation of China (Grant No. 11204035). The numerical calculations were performed at the Center of Advanced Computing and Data Systems at the University of Houston.
\end{acknowledgments}\index{}

\begin{appendix}

\section{realization of type-II Weyl node}

In this section, we perform an analysis on the Weyl nodes that are supported in the present system defined by Fig. 1 and Eq.(3) of the main text. In particular, as Fig.3 has shown the transition from Type-I Weyl node to type-II Weyl node, it is interesting to investigate the possibility of realizing type-II Weyl nodes in other cases. From the analysis presented below, the type-II Weyl nodes found in Fig.3 for $\eta(\mathbf{k})=\sin k_{x}\sin k_{y}$ are the only cases where type-II Weyl nodes can be realized in the proposed system. We also point out the possibility of realizing semi-Weyl nodes.

The low-energy effective model for a single Weyl node can be written generally in terms of the relative momenta $\mathbf{q}$ as \cite{soluyanov15}
\begin{eqnarray}
h_{\nu}(\mathbf{q})&=&\sum\limits_{i=1}^{2}v_{\nu i}^{0}q_{i}\tau_{0}+\sum\limits_{i=1}^{2}\sum\limits_{j=1}^{3}q_{i}A^{\nu}_{ij}\tau_{j}  \notag \\
&=&\sum\limits_{i=1}^{2}v^{0}_{\nu i}q_{i}\tau_{0}+\sum\limits_{j=1}^{3}(\sum\limits_{i=1}^{2}q_{i}A^{\nu}_{ij})\tau_{j}   \notag \\
&=&v_{\nu0}(\mathbf{q})\tau_{0}+\sum\limits_{j=1}^{3}v_{\nu j}(\mathbf{q})\tau_{j},
\end{eqnarray}
where $\nu$ is a label for the specific Weyl node under consideration, $\tau_{0}$ is the second-order unit matrix. The two eigenvalues of the model are
\begin{equation}
\varepsilon_{\nu\pm}(\mathbf{q})=v_{\nu0}(\mathbf{q})\pm\sqrt{\sum\limits_{j=1}^{3}v^{2}_{\nu j}(\mathbf{q})}=v_{\nu0}(\mathbf{q})\pm v_{\nu}(\mathbf{q}).
\end{equation}
$h(\mathbf{q})$ describes a type-II Weyl node if and only if there exists a direction $\hat{\mathbf{q}}$ along which $|v_{\nu0}(\mathbf{q})|>v_{\nu}(\mathbf{q})$. Otherwise, the Weyl node is type-I.

For the heterostructure with $s$-wave superconductor, the Weyl superconductivity phase ($s$WSC) appears only when the magnetization lies along in-plane directions ($\theta=\pi/2$ and $3\pi/2$). From the low-energy effective model of the Weyl nodes for $\theta=\pi/2$ \cite{hao16}, the Weyl nodes can become type-II only when the strength of the RSOC ($\lambda$) is of the same order of magnitude as the hopping amplitude $t$. Since this is unrealistic, the Weyl nodes in the $s$WSC can only be type-I.

For the $d$WSC with $\eta(\mathbf{k})=\cos k_{x}-\cos k_{y}$, the two Weyl nodes emerging for in-plane magnetization are qualitatively the same as the Weyl nodes for $s$WSC. Following the same analysis as above, they can only be type-I for realistic model parameters. The remaining four Weyl nodes exist for all $\theta$ angles. Their positions in the 2D BZ are nearly $\theta$-independent and are $\mathbf{k}_{\alpha\beta}=(\alpha,\beta)k_{x0}$ ($\alpha=\pm$, $\beta=\pm$). $k_{x0}$ is defined by Eq.(10). For a general $\theta$, we have
\begin{equation}
v_{\alpha\beta0}(\mathbf{q})=\frac{\lambda m_{x}}{m}(\beta\sin k_{x0}+q_{y}\cos k_{x0}),
\end{equation}
\begin{eqnarray}
v_{\alpha\beta1}(\mathbf{q})&=&(\alpha q_{x}-\beta q_{y})\frac{-2\beta\lambda\Delta_{0}\sin^{2}k_{x0}}{m(m+m_{z})}  \notag \\
&& \cdot(m_{z}-m+\frac{2m^{2}_{x}-\lambda^{2}\sin^{2}k_{x0}}{m}),
\end{eqnarray}
\begin{eqnarray}
v_{\alpha\beta2}(\mathbf{q})&=&(\alpha q_{x}-\beta q_{y})\frac{-2\alpha\lambda\Delta_{0}\sin^{2}k_{x0}}{m(m+m_{z})}   \notag \\
&& (m_{z}-m-\frac{\lambda^{2}\sin^{2}k_{x0}}{m}),
\end{eqnarray}
and
\begin{equation}
v_{\alpha\beta3}(\mathbf{q})=(\alpha q_{x}+\beta q_{y})(2t\sin k_{x0}-\frac{\lambda^{2}\sin 2k_{x0}}{2m}).
\end{equation}
The constant term in $v_{\alpha\beta0}(\mathbf{q})$ gives the energy of the Weyl node but is irrelevant to the type of the Weyl node. $v_{\alpha\beta0}(\mathbf{q})$ only gives a velocity along the $q_{y}$ direction, of the order no larger than $\lambda$. $v_{\alpha\beta3}(\mathbf{q})$, on the other hand gives a velocity of the same order as $t$ along $q_{y}$. As a result, the four Weyl nodes $\mathbf{k}_{\alpha\beta}$ ($\alpha=\pm$, $\beta=\pm$) are type-I for realistic model parameters and all $\theta$ values.

For the $d$WSC with $\eta(\mathbf{k})=\sin k_{x}\sin k_{y}$, there are always four Weyl nodes located approximately at $\mathbf{k}_{\alpha\beta}=\alpha(\frac{1+\beta}{2},\frac{1-\beta}{2})k_{xc}$ ($\alpha=\pm$, $\beta=\pm$). To simplify notation, we introduce the following abbreviations: $s_{m\beta}=\sin\frac{1-\beta}{2}k_{xc}$, $c_{m\beta}=\cos\frac{1-\beta}{2}k_{xc}$, $s_{p\beta}=\sin\frac{1+\beta}{2}k_{xc}$, $c_{p\beta}=\sin\frac{1+\beta}{2}k_{xc}$, $s_{2m\beta}=2s_{m\beta}c_{m\beta}$, $s_{2p\beta}=2s_{p\beta}c_{p\beta}$. For a general $\theta$, the effective low-energy model for the Weyl nodes are
\begin{equation}
v_{\alpha\beta0}(\mathbf{q})=\frac{\lambda m_{x}}{m}(\alpha s_{m\beta}+q_{y}c_{m\beta}),
\end{equation}
\begin{eqnarray}
v_{\alpha\beta1}(\mathbf{q})&=&(s_{m\beta}c_{p\beta}q_{x}+s_{p\beta}c_{m\beta}q_{y})\frac{2\lambda\Delta_{0}s_{m\beta}}{m(m+m_{z})}  \notag \\
&& \cdot(m_{z}-m+\frac{4m^{2}_{x}-\lambda^{2}(s^{2}_{p\beta}+s^{2}_{m\beta})}{2m}),
\end{eqnarray}
\begin{eqnarray}
v_{\alpha\beta2}(\mathbf{q})&=&(s_{m\beta}c_{p\beta}q_{x}+s_{p\beta}c_{m\beta}q_{y})\frac{2\lambda\Delta_{0}s_{p\beta}}{m(m+m_{z})}  \notag \\
&& \cdot(m_{z}-m-\frac{\lambda^{2}(s^{2}_{p\beta}+s^{2}_{m\beta})}{2m}),
\end{eqnarray}
and
\begin{equation}
v_{\alpha\beta3}(\mathbf{q})=\alpha q_{x}(2ts_{p\beta}-\frac{\lambda^{2}s_{2p\beta}}{2m})+\alpha q_{y}(2ts_{m\beta}-\frac{\lambda^{2}s_{2m\beta}}{2m}).
\end{equation}
Again, because $v_{\alpha\beta0}(\mathbf{q})$ only gives the quasiparticle a velocity along $q_{y}$, only if the velocity along $q_{y}$ from $v_{\alpha\beta}(\mathbf{q})=\sqrt{\sum^{3}_{i=1}v^{2}_{\alpha\beta i}(\mathbf{q})}$ is smaller than that from $v_{\alpha\beta0}(\mathbf{q})$, can the Weyl node become type-II. Note that $s_{m+}=s_{p-}=0$, the effective model for the two Weyl nodes $\mathbf{k}_{\alpha-}=\alpha(0,1)k_{xc}$ becomes
\begin{equation}
v_{\alpha-0}(\mathbf{q})=\frac{\lambda m_{x}}{m}(\alpha\sin k_{xc}+q_{y}\cos k_{xc}),
\end{equation}
\begin{equation}
v_{\alpha-1}(\mathbf{q})=q_{x}\frac{2\lambda\Delta_{0}\sin^{2}k_{xc}}{m(m+m_{z})}(m_{z}-m-\frac{\lambda^{2}\sin^{2}k_{xc}-4m^{2}_{x}}{2m}),
\end{equation}
\begin{equation}
v_{\alpha-3}(\mathbf{q})=\alpha q_{y}(2t\sin k_{xc}-\frac{\lambda^{2}\sin2k_{xc}}{2m}),
\end{equation}
and $v_{\alpha-2}(\mathbf{q})=0$. From the above effective model, the two Weyl nodes at $\mathbf{k}_{\alpha-}=\alpha(0,1)k_{xc}$ can only be type-II when $\lambda$ is of the same order as $t$, which is unrealistic in the present setting. An interesting critical case for these two nodes is when the coefficient of $q_{x}$ in Eq.(A12) becomes zero. When this happens, the linear in $q_{x}$ term vanishes, we have to add the term quadratic in $q_{x}$. The nodes thus become linear in $q_{y}$ but quadratic in $q_{x}$, and is thus semi-Weyl \cite{pardo09,banerjee09}. In a HM with large magnetic anisotropy energy, this critical point corresponding to a general $\theta$ angle should be very difficult to attain. If the magnetic anisotropy energy is small so that we can tune $\theta$ through the full $2\pi$ range, it is still possible to turn the two Weyl nodes at $\mathbf{k}_{\alpha-}=\alpha(0,1)k_{xc}$ to be semi-Weyl.


The effective model for the other pair of Weyl nodes, $\mathbf{k}_{\alpha+}=\alpha(1,0)k_{xc}$, are
\begin{equation}
v_{\alpha+0}(\mathbf{q})=\frac{\lambda m_{x}}{m}q_{y},
\end{equation}
\begin{equation}
v_{\alpha+2}(\mathbf{q})=q_{y}\frac{2\lambda\Delta_{0}\sin^{2}k_{xc}}{m(m+m_{z})}(m_{z}-m-\frac{\lambda^{2}\sin^{2}k_{xc}}{2m}),
\end{equation}
\begin{equation}
v_{\alpha+3}(\mathbf{q})=\alpha q_{x}(2t\sin k_{xc}-\frac{\lambda^{2}\sin2k_{xc}}{2m}),
\end{equation}
and $v_{\alpha+1}(\mathbf{q})=0$. For experimentally relevant parameters satisfying $|\lambda/t|\ll1$, $|\Delta_{0}/t|\ll1$, and $|m/t|\sim 1$, the type-II Weyl node is clearly achievable. In the limiting cases of $|\lambda|/m\ll1$ and $|\Delta_{0}|/m\ll1$, we can also get an analytical expression for the critical angles marking the transition between type-I and type-II Weyl nodes,
\begin{equation}
|\sin\theta_{C}|\simeq\frac{|\Delta_{0}|}{2m}(\frac{\lambda}{m})^{2}\sin^{4}k_{xc}.
\end{equation}

Overall, we have verified both numerically and analytically that a transition between type-I and type-II Weyl nodes can be achieved in our proposed experimental configuration, related to the pair of Weyl nodes lying along $(k_{x},0)$ for $\eta(\mathbf{k})=\sin k_{x}\sin k_{y}$. Finally, we reemphasize that the two cases, $\eta(\mathbf{k})=\cos k_{x}-\cos k_{y}$ and $\eta(\mathbf{k})=\sin k_{x}\sin k_{y}$, can be realized by growing the same $d$-wave superconductor on the HM thin film along two inequivalent special directions.

\end{appendix}



\begin{references}

\bibitem{kane05} C. L. Kane and E. J. Mele, Phys. Rev. Lett. \textbf{95}, 146802 (2005); \emph{ibid} \textbf{95}, 226801 (2005).


\bibitem{bernevig06} B. Andrei Bernevig, Taylor L. Hughes, Shou-Cheng Zhang, Science \textbf{314}, 1757 (2006).


\bibitem{fu07} Liang Fu, C. L. Kane, and E. J. Mele, Phys. Rev. Lett. \textbf{98}, 106803 (2007).



\bibitem{moore07} J. E. Moore and L. Balents, Phys. Rev. B \textbf{75}, 121306(R) (2007).


\bibitem{roy09} Rahul Roy, Phys. Rev. B \textbf{79}, 195322 (2009).



\bibitem{zhang09} Haijun Zhang, Chao-Xing Liu, Xiao-Liang Qi, Xi Dai, Zhong Fang, and Shou-Cheng Zhang, Nature Phys. \textbf{5}, 438 (2009).



\bibitem{konig07} Markus K\"{o}nig, Steffen Wiedmann, Christoph Br\"{u}ne, Andreas Roth, Hartmut Buhmann, Laurens W. Molenkamp, Xiao-Liang Qi, and Shou-Cheng Zhang, Science \textbf{318}, 766 (2007).



\bibitem{hsieh08} D. Hsieh, D. Qian, L. Wray, Y. Xia, Y. S. Hor, R. J. Cava, and M. Z. Hasan, Nature \textbf{452}, 970 (2008).



\bibitem{xia09} Y. Xia, D. Qian, D. Hsieh, L. Wray, A. Pal, H. Lin, A. Bansil, D. Grauer, Y. S. Hor, R. J. Cava, and M. Z. Hasan, Nature Phys. \textbf{5}, 398 (2009).



\bibitem{schnyder08} Andreas P. Schnyder, Shinsei Ryu, Akira Furusaki, and Andreas W. W. Ludwig, Phys. Rev. B \textbf{78}, 195125 (2008).



\bibitem{kitaev09} Alexei Kitaev, AIP Conf. Proc. \textbf{1134}, 22 (2009).



\bibitem{hasan10} M. Z. Hasan and C. L. Kane, Rev. Mod. Phys. \textbf{82}, 3045 (2010).


\bibitem{qi11} Xiao-Liang Qi and Shou-Cheng Zhang, Rev. Mod. Phys. \textbf{83}, 1057 (2011).



\bibitem{chenx13} Xie Chen, Zheng-Cheng Gu, Zheng-Xin Liu, and Xiao-Gang Wen, Phys. Rev. B \textbf{87}, 155114 (2013).



\bibitem{mesaros13} Andrej Mesaros and Ying Ran, Phys. Rev. B \textbf{87}, 155115 (2013).


\bibitem{lu16} Yuan-Ming Lu and Ashvin Vishwanath, Phys. Rev. B \textbf{93}, 155121 (2016).


\bibitem{wan11} Xiangang Wan, Ari M. Turner, Ashvin Vishwanath, and Sergey Y. Savrasov, Phys. Rev. B \textbf{83}, 205101 (2011).


\bibitem{burkov11} A. A. Burkov and Leon Balents, Phys. Rev. Lett. \textbf{107}, 127205 (2011).


\bibitem{murakami07} Shuichi Murakami, New J. Phys. \textbf{9}, 356 (2007).


\bibitem{wehling14} T. O. Wehling, A. M. Black-Schaffer, and A. V. Balatsky, Adv. Phys. \textbf{63}, 1 (2014).


\bibitem{yang11} Kai-Yu Yang, Yuan-Ming Lu, and Ying Ran, Phys. Rev. B \textbf{84}, 075129 (2011).


\bibitem{hosur12} Pavan Hosur, S. A. Parameswaran, and Ashvin Vishwanath, Phys. Rev. Lett. \textbf{108}, 046602 (2012).


\bibitem{zyuzin12} A. A. Zyuzin and A. A. Burkov, Phys. Rev. B \textbf{86}, 115133 (2012).


\bibitem{liu13} Chao-Xing Liu, Peng Ye, and Xiao-Liang Qi, Phys. Rev. B \textbf{87}, 235306 (2013).



\bibitem{hosur13} Pavan Hosur and Xiaoliang Qi, C. R. Phys. \textbf{14}, 857 (2013).



\bibitem{young12} S. M. Young, S. Zaheer, J. C.Y. Teo, C. L. Kane, E. J. Mele, and A. M. Rappe, Phys. Rev. Lett. \textbf{108}, 140405 (2012).


\bibitem{young15} Steve M. Young and Charles L. Kane, Phys. Rev. Lett. \textbf{115}, 126803 (2015).



\bibitem{wang12} Zhijun Wang, Yan Sun, Xing-Qiu Chen, Cesare Franchini, Gang Xu, Hongming Weng, Xi Dai, and Zhong Fang, Phys. Rev. B \textbf{85}, 195320 (2012).


\bibitem{wang13} Zhijun Wang, Hongming Weng, Quansheng Wu, Xi Dai, and Zhong Fang, Phys. Rev. B \textbf{88}, 125427 (2013).


\bibitem{yang14nc} Bohm-Jung Yang and Naoto Nagaosa, Nature Commun. \textbf{5}, 4898 (2014).


\bibitem{klinovaja16} Jelena Klinovaja, Peter Stano, and Daniel Loss, Phys. Rev. Lett. \textbf{116}, 176401 (2016).


\bibitem{cho12} Gil Young Cho, Jens H. Bardarson, Yuan-Ming Lu, and Joel E. Moore, Phys. Rev. B \textbf{86}, 214514 (2012).


\bibitem{meng12} Tobias Meng and Leon Balents, Phys. Rev. B \textbf{86}, 054504 (2012).



\bibitem{yang14} Shengyuan A. Yang, Hui Pan, and Fan Zhang, Phys. Rev. Lett. \textbf{113}, 046401 (2014).


\bibitem{sato16} Masatoshi Sato and Satoshi Fujimoto, J. Phys. Soc. Jpn. \textbf{85}, 072001 (2016).



\bibitem{neumann29} J. von Neumann and E. Wigner, Z. Phys. \textbf{30}, 467 (1929).



\bibitem{blount85} E. I. Blount, Phys. Rev. B \textbf{32}, 2935 (1985).



\bibitem{balents11} Leon Balents, Physics \textbf{4}, 36 (2011).



\bibitem{hou13} J. M. Hou, Phys. Rev. Lett. \textbf{111}, 130403 (2013).



\bibitem{lee09} Patrick A. Lee, arXiv:0907.2681.



\bibitem{chung11} Suk Bum Chung, Hai-Jun Zhang, Xiao-Liang Qi, and Shou-Cheng Zhang, Phys. Rev. B \textbf{84}, 060510(R) (2011).


\bibitem{gorkov01} Lev P. Gor’kov and Emmanuel I. Rashba, Phys. Rev. Lett. \textbf{87}, 037004 (2001).



\bibitem{fu08} Liang Fu and C. L. Kane, Phys. Rev. Lett. \textbf{100}, 096407 (2008).


\bibitem{zhang08} C. Zhang, S. Tewari, R. M. Lutchyn, and S. D. Sarma, Phys. Rev. Lett. \textbf{101}, 160401 (2008).



\bibitem{sato09} M. Sato, Y. Takahashi, and S. Fujimoto, Phys. Rev. Lett. \textbf{103}, 020401 (2009); M. Sato and S. Fujimoto, Phys. Rev. B \textbf{79}, 094504 (2009).



\bibitem{sau10} J. D. Sau, R. M. Lutchyn, S. Tewari, and S. Das Sarma, Phys. Rev. Lett. \textbf{104}, 040502 (2010).


\bibitem{alicea10} Jason Alicea, Phys. Rev. B \textbf{81}, 125318 (2010).


\bibitem{hao16} Lei Hao and C. S. Ting, Phys. Rev. B \textbf{94}, 134513 (2016).



\bibitem{sato10} Masatoshi Sato and Satoshi Fujimoto, Phys. Rev. Lett. \textbf{105}, 217001 (2010).


\bibitem{wongetal} Chris L. M. Wong, Jie Liu, K. T. Law, and Patrick A. Lee, Phys. Rev. B \textbf{88}, 060504(R) (2013); Florian Loder, Arno P. Kampf, Thilo Kopp, Sci. Rep. \textbf{5}, 15302 (2015); Akito Daido and Youichi Yanase, Phys. Rev. B \textbf{94}, 054519 (2016); Akito Daido and Youichi Yanase, arXiv:1611.09492v1.



\bibitem{soluyanov15} Alexey A. Soluyanov, Dominik Gresch, Zhijun Wang, QuanSheng Wu, Matthias Troyer, Xi Dai, and B. Andrei Bernevig, Nature (London) \textbf{527}, 495 (2015).




\bibitem{wang16} Zhijun Wang, Dominik Gresch, Alexey A. Soluyanov, Weiwei Xie, S. Kushwaha, Xi Dai, Matthias Troyer, Robert J. Cava, and B. Andrei Bernevig, Phys. Rev. Lett. \textbf{117}, 056805 (2016).




\bibitem{chang16} Guoqing Chang, Su-Yang Xu, Daniel S. Sanchez, Shin-Ming Huang, Chi-Cheng Lee, Tay-Rong Chang, Guang Bian, Hao Zheng, Ilya Belopolski, Nasser Alidoust, Horng-Tay Jeng, Arun Bansil, Hsin Lin, and M. Zahid Hasan, Sci. Adv. \textbf{2}, e1600295 (2016).



\bibitem{xu15} Yong Xu, Fan Zhang, and Chuanwei Zhang, Phys. Rev. Lett. \textbf{115}, 265304 (2015).



\bibitem{lifshitz} I. M. Lifshitz, Sov. Phys. JETP \textbf{11}, 1130 (1960).



\bibitem{bcs} J. Bardeen, L. N. Cooper, and J. R. Schrieffer, Phys. Rev. \textbf{108}, 1175 (1957).



\bibitem{ff} Peter Fulde and Richard A. Ferell, Phys. Rev. \textbf{135}, A550 (1964).


\bibitem{lo} A. I. Larkin and Y. N. Ovchinnikov, Zh. Eksp. Teor. Fiz. \textbf{47}, 1136 (1964) [Sov. Phys. JETP \textbf{20}, 762 (1965)].



\bibitem{liu03} W. Vincent Liu and Frank Wilczek, Phys. Rev. Lett. \textbf{90}, 047002 (2003).


\bibitem{gubankova03} Elena Gubankova, W. Vincent Liu, and Frank Wilczek, Phys. Rev. Lett. \textbf{91}, 032001 (2003).


\bibitem{bedaque03} Paulo F. Bedaque, Heron Caldas, and Gautam Rupak, Phys. Rev. Lett. \textbf{91}, 247002 (2003).



\bibitem{sarma63} G. Sarma, J. Phys. Chem. Solids \textbf{24}, 1029 (1963).


\bibitem{arpes} Andrea Damascelli, Zahid Hussain, and Zhi-Xun Shen, Rev. Mod. Phys. \textbf{75}, 473 (2003).


\bibitem{sts} Chia-Ren Hu, Phys. Rev. Lett. \textbf{72}, 1526 (1994); Yukio Tanaka and Satoshi Kashiwaya, Phys. Rev. Lett. \textbf{74}, 3451 (1995); Satoshi Kashiwaya and Yukio Tanaka, Rep. Prog. Phys. \textbf{63}, 1641 (2000); Guy Deutscher, Rev. Mod. Phys. \textbf{77}, 109 (2005).


\bibitem{hao11} Lei Hao and T. K. Lee, Phys. Rev. B \textbf{83}, 134516 (2011); Lei Hao, Gui-Ling Wang, Ting-Kuo Lee, Jun Wang, Wei-Feng Tsai, and Yong-Hong Yang, Phys. Rev. B \textbf{89}, 214505 (2014).




\bibitem{volovik16} G. E. Volovik, arXiv:1606.08318v6.




\bibitem{buzdin90} A. I. Buzdin and M. Y. Kuprianov, Pis'ma Zh. Eksp. Teor. Fiz. \textbf{52}, 1089 (1990) [JETP Lett. \textbf{52}, 487 (1990)].




\bibitem{radovic91} Zoran Radovi\'{c}, Marko Ledvij, Ljiljana Dobrosavljevi\'{c}-Gruji\'{c}, A. I. Buzdin, and John R. Clem, Phys. Rev. B \textbf{44}, 759     (1991).



\bibitem{demler97} E. A. Demler, G. B. Arnold, and M. R. Beasley, Phys. Rev. B \textbf{55}, 15174 (1997).



\bibitem{eschrig03} M. Eschrig, J. Kopu, J. C. Cuevas, and Gerd Sch\"{o}n, Phys. Rev. Lett. \textbf{90}, 137003 (2003).




\bibitem{buzdin05} A. I. Buzdin, Rev. Mod. Phys. \textbf{77}, 935 (2005).




\bibitem{lu15} B. Lu, K. Yada, M. Sato, and Y. Tanaka, Phys. Rev. Lett. \textbf{114}, 096804 (2015).



\bibitem{ghosh10} P. Ghosh, J. D. Sau, S. Tewari, and S. D. Sarma, Phys. Rev. B \textbf{82}, 184525 (2010).



\bibitem{sato10b} Masatoshi Sato, Phys. Rev. B \textbf{81}, 220504(R) (2010).


\bibitem{zak89} J. Zak, Phys. Rev. Lett. \textbf{62}, 2747 (1989).



\bibitem{delplace11} P. Delplace, D. Ullmo, and G. Montambaux, Phys. Rev. B \textbf{84}, 195452 (2011).



\bibitem{sato11} Masatoshi Sato, Yukio Tanaka, Keiji Yada, and Takehito Yokoyama, Phys. Rev. B \textbf{83}, 224511 (2011); Keiji Yada, Masatoshi Sato, Yukio Tanaka, and Takehito Yokoyama, Phys. Rev. B \textbf{83}, 064505 (2011); Akihiro Ii, Keiji Yada, Masatoshi Sato, and Yukio Tanaka, Phys. Rev. B \textbf{83}, 224524 (2011).



\bibitem{zhu00} Jian-Xin Zhu and C. S. Ting, Phys. Rev. B \textbf{61}, 1456 (2000); Zhi Ping Niu and D. Y. Xing, Phys. Rev. Lett. \textbf{98}, 057005 (2007).



\bibitem{pardo09} Victor Pardo and Warren E. Pickett, Phys. Rev. Lett. \textbf{102}, 166803 (2009).



\bibitem{banerjee09} S. Banerjee, R. R. P. Singh, V. Pardo, and W. E. Pickett, Phys. Rev. Lett. \textbf{103}, 016402 (2009).






\end{references}
\end{document}